\documentclass[aps,prd,twocolumn,nofootinbib,groupedaddress]{revtex4-1}
\usepackage{aas_macros}
\usepackage{amsmath}
\usepackage{amsfonts}
\usepackage{bm}
\usepackage{graphicx}
\usepackage{xspace}

\usepackage{xcolor}

\newcommand\SkipCrapAtEnd[1]{}

\newcommand\qmstateproduct[2]{\left\langle#1|#2\right\rangle}

\definecolor{todocomment}{HTML}{D62728}
\definecolor{dwcomment}{HTML}{1F77B4}
\definecolor{jlcomment}{HTML}{2CA02C}
\definecolor{rocomment}{HTML}{9467BD}

\newcommand{\IMRPD}{\textsc{IMRPhenomD}\xspace}

\newcommand\RIFT{RIFT}

\newcommand\E[1]{{\left\langle #1 \right \rangle}}
\newcommand\unit[1]{{\rm #1}}
\newcommand\rate{\mathcal{R}}

\newcommand\chieff{\chi_{\text{eff}}}
\newcommand\mchirp{\mathcal{M}_{\text{c}}}
\newcommand\mc{\mchirp}

\newcommand\param{\lambda}
\newcommand\Param{\Lambda}

\newcommand\Vc{V_{\mathrm{c}}}

\newcommand\mmin{m_{\mathrm{min}}}
\newcommand\mmax{m_{\mathrm{max}}}
\newcommand\Mmax{M_{\mathrm{max}}}

\newcommand\Msun{\mathrm{M}_{\odot}}

\newcommand\cqg{CQG}

\begin{document}

\preprint{P1800107-v4}

\title{Reconstructing phenomenological distributions of compact binaries via gravitational wave observations}%

\author{Daniel Wysocki}
\email{dw2081@rit.edu}
\affiliation{Rochester Institute of Technology, Rochester, New York 14623, USA}

\author{Jacob Lange}
\affiliation{Rochester Institute of Technology, Rochester, New York 14623, USA}

\author{Richard O'Shaughnessy}
\affiliation{Rochester Institute of Technology, Rochester, New York 14623, USA}

\date{\today}

\begin{abstract}
Gravitational wave (GW) measurements will provide insight into the population of coalescing compact binaries throughout the universe.
We describe and demonstrate a flexible parametric method to  infer the event rate as a function of compact binary
parameters, accounting for Poisson error and selection biases.    
Using  synthetic data based on projections for LIGO and Virgo's third observing  run (O3), we discuss how well GW measurements could constrain the mass and spin distribution of
coalescing neutron stars and black holes (BHs) in the near future, within the context of several phenomenological models
described in this work.  We demonstrate that only a few tens of events can enable astrophysically significant
constraints on the spin magnitude and orientation distribution of BHs in merging binaries.
We discuss how astrophysical priors or other measurements can inform the interpretation of future measurements.
 Using publicly available results, we estimate the event rate versus mass for binary black holes (BBHs).  To connect to
 previously published work, we provide estimates including  reported O2 BBH candidates,
 making several unwarranted but simplifying assumptions for the sensitivity of the network and completeness of the reported
 set of events.   Consistent with prior
 work, we find BHs in binaries likely have low natal spin.  With available results and a population favoring low  spin, we
 cannot presently constrain the typical misalignments of the binary black hole population.  
All of the tools described in this work are publicly available and ready-to-use to interpret real or synthetic LIGO
data, and to synthesize projected data from future observing
runs.\footnote{\url{https://git.ligo.org/daniel.wysocki/bayesian-parametric-population-models/} and \url{https://git.ligo.org/daniel.wysocki/synthetic-PE-posteriors}}
\end{abstract}

\maketitle

\section{
  \label{sec:intro}
  Introduction
}

The Advanced Laser Interferometer Gravitational Wave Observatory (LIGO) ~\cite{2015CQGra..32g4001T} and  Virgo
\cite{gw-detectors-Virgo-original-preferred,gw-detectors-Virgo-new}
detectors have
and will continue to discover gravitational waves (GW) from coalescing binary black holes (BBHs) and neutron stars.  Several
tens of binary black holes and potentially neutron stars are expected to be seen in O3, LIGO's next observing run, alone; and several hundreds more detections are expected over the next five years  \cite{RatesPaper,LIGO-O1-BBH}.
Already, the properties  of the sources responsible -- the inferred event rates, masses, and spins -- have confronted
other observations of black holes' masses and spins \cite{LIGO-O1-BBH}, challenged
previous  formation scenarios \cite{AstroPaper,LIGO-O1-BBH}, and inspired  new models
\cite{2016MNRAS.458.2634M,popsyn-ChemHomogeneous-Marchant2016,2016ApJ...824L...8R,2016PhRvL.116t1301B} and insights
\cite{2016MNRAS.462..844K,2016MNRAS.463L..31L} into the evolution of massive stars and the observationally accessible
gravitational waves they emit \cite{2016MNRAS.461.3877D,StochasticPaper}.    Over the next
several years, our understanding of the lives and deaths of massive stars over cosmic time will be transformed by the identification and
interpretation of the population(s) responsible for coalescing binaries \cite{AstroPaper,2018arXiv180605195B,2018PhRvD..97d3014W},  because  measurements will enable  robust tests to distinguish between
formation scenarios \cite{2010CQGra..27k4007M} with present \cite{gwastro-popsynVclusters-Rodriguez2016} and future
instruments \cite{2016ApJ...830L..18B,2016PhRvD..94f4020N}. %

During the first few years of discovery,  substantial theoretical modeling challenges and the rapid
pace of events suggest that GW observations could soon outpace theory.  In this work, we introduce a flexible, concrete, and
production-ready approach  to infer compact binary merger rate and compact binary distribution, in the context of an
(arbitrary) parametrized phenomenological model.  We extend or employ previously proposed
models \cite{2017ApJ...851L..25F,2018ApJ...856..173T}.   We are motivated by how constraints on these phenomenological models enable us
to address  broad  astrophysical questions---the mass and spin distribution of neutron stars and black holes, as
imparted at their birth;  the dominant formation mechanism for compact binaries, such as the role of dynamical  versus
isolated formation channels for binary black holes.  To that end, we provide concrete demonstrations of how
a few GW measurements will provide insights that enable sharp discrimination between proposed astrophysical
alternatives, or measurements of their parameters.   We use simple phenomenological arguments
and calculations to characterize the information that these first few hundred observations should provide.   
Conversely, we provide simple approaches to  extend our phenomenological approach in sophistication and complexity as
several thousand compact binary mergers provide sharp constraints on their underlying properties. 
This approach complements inferences that work within a concrete model family as explored in other proof-of-concept investigations
(see, e.g., \cite{2010CQGra..27k4007M,2013PhRvD..88h4061O,2015ApJ...810...58S,gwastro-EventPopsynPaper-2016,2017ApJ...846...82Z,2018MNRAS.tmp..873B,2017PhRvD..96f4025M,2018PhRvD..97d3014W} and
references therein).

GW measurements probe only a selection-biased part of the compact binary distribution.  Previously reported estimates of the overall
compact binary event rate rely  on extrapolation away from the observed population, using some fixed model
for the compact binary mass distribution \cite{LIGO-O1-BBH}. 
In fact, the  compact binary mass distribution and inferred event rate are strongly coupled.
This paper provides the first  self-consistent approach to infer  both the compact binary event rate and
parameter distribution; then it describes and explains the expected correlation in an accessible way.

Several recent studies have explored how well GW measurements can constrain the mass and spin distribution of binary
black holes %
 \cite{2013PhRvD..88h4061O,WysockiThesis,2017MNRAS.465.3254M,2017PhRvD..95j3010K,2017PhRvD..96b3012T,2017Natur.548..426F,2017ApJ...851L..25F,2017PhRvD..95l4046G,2017ApJ...840L..24F,2017MNRAS.471.2801S,LIGO-O1-BBH,2017PhRvD..95f4053V,2018ApJ...854L...9F}.
Our approach is novel insofar as it reconstructs both the strongly correlated event rate and the parameter distribution,
making our method a robust tool to assess astrophysical formation scenarios.   In our modeling, we focus on measuring
the black hole (BH) spin magnitude and misalignment distribution, as a method to probe the formation scenarios for binary BHs.
As first described in \cite{2010CQGra..27k4007M}, GW provide a unique opportunity to distinguish between
isolated and dynamic formation mechanisms: measurements of the spin properties of
the BHs~\cite{AstroPaper,gwastro-popsynVclusters-Rodriguez2016,gwastro-PE-Salvo-EvidenceForAlignment-2015,2017MNRAS.471.2801S,gwastro-ConstrainChannels-BoxingDayKicks-Me2017, 2017PhRvD..96b3012T}.
The presence of a component of the BH spins in the plane of the orbit leads to precession of that plane.  If suitably massive and significantly spinning, such binaries will strongly precess within the LIGO sensitive band.
If BBHs are the end points of isolated binary star systems, they would be expected to contain BHs with spins
preferentially aligned with the orbital angular momentum~\cite{2000ApJ...541..319K,gwastro-ConstrainChannels-BoxingDayKicks-Me2017}, and therefore rarely  be strongly precessing.
If, however, BBHs predominantly form as a result of gravitational interactions inside dense populations of stellar systems, the relative orientations of the BH spins with their orbits will be random, and some gravitational wave signals may be very strongly precessing.
At this early stage, observations cannot firmly distinguish between these two scenarios, or more broadly other possible BBH formation mechanisms~\cite{AstroPaper}.
These include the evolution of isolated pairs of
stars~\cite{gwastro-EventPopsynPaper-2016,popsyn-LowMetallicityImpact-Chris2010,popsyn-LIGO-SFR-2008,2012ApJ...759...52D,2016MNRAS.458.2634M,popsyn-ChemHomogeneous-Marchant2016},
dynamic binary formation in dense clusters~\cite{2016ApJ...824L...8R}, and pairs of primordial black holes BHs~\cite{2016PhRvL.116t1301B}; see, e.g.,~\cite{AstroPaper} and references therein.  
Loosely speaking, however, the isolated evolution and globular cluster formation scenarios are the most well-developed
and verifiable using independent observational constraints.  
More broadly, precise measurements of their properties will provide unique clues into how BHs and massive stars evolve~\cite{gwastro-PE-Salvo-EvidenceForAlignment-2015,2017MNRAS.471.2801S,gwastro-popsynVclusters-Rodriguez2016,2017Natur.548..426F,gwastro-ConstrainChannels-KickRatePaper-2017,2016ApJ...830L..18B,2016PhRvD..94f4020N}.

This paper is organized as follows.
In Sec. \ref{sec:method} we describe our techniques to infer compact binary populations, building upon inferences
about parameters of individual events.    Unlike prior work, we simultaneously reconstruct the event rate, mass
distribution, and spin (vector) distribution. 
In Sec. \ref{sec:tests1}, we demonstrate our our population inference strategy with two examples.  In
the first, we perform a full end-to-end analysis of a synthetic
GW data generated from a synthetic population of astrophysically distributed sources.    In the second, using a tool to
mimic how well we could constrain parameters of a candidate GW signal, we perform a large-scale investigation into how
well GW measurements could constrain the mass and spin distribution of binary black holes.   We find that the mass and
spin distribution can be tightly constrained with only a few tens of events.  By virtue of explicitly
exploiting only some of the available information, our estimates are necessarily conservative. 
In Sec. \ref{sec:RealEventAnalysis}, we apply our method to the currently reported binary black hole population.  For
simplicity, assuming the reported events to date represent a fair sample of 
the results of LIGO's first two observing runs (O1 and O2), we corroborate
previous results, finding black hole
spins are likely small and that the black hole mass spectrum may have an upper bound.   Due to small BH spins, except
for GW151226, we can extract no information about typical BBH spin-orbit misalignments.   
We emphasize our demonstration uses a nonfinal sample for LIGO's O2 survey: depending on that survey's results, applying our methods to final O2 results
could produce  substantially different astrophysical conclusions.  
In Sec. \ref{sec:discussion} we briefly discuss the accuracy to which population parameters can be determined, and
the surprisingly significant role of waveform systematics in the near future.  
After summarizing our conclusions in Sec. \ref{sec:conclusions}, we supply three  appendixes. 
In Appendix \ref{ap:mock}, we describe a robust, extensible procedure for generating synthetic posterior distributions
for proposed GW events.  This open-source procedure could be widely used to assess the viability of GW measurements to
distinguish between proposed astrophysical channels.  A subsequent short Appendix \ref{ap:mockPop} describes how to
generate synthetic populations of selection-biased GW sources using this procedure. 
Next, in Appendix \ref{ap:UnderstandLimits}, following on and extending previous work, we use toy models for both the measurement
process and source population to illustrate  how well GW observations will constrain the mass and spin distribution of
compact binaries, likely providing robust insights into compact object formation (e.g., BH natal spins and maximum
masses) and binary formation mechanisms (e.g., dynamical over isolated). %

\section{Method}
\label{sec:method}

\subsection{Characterizing and inferring parameters of individual binary black holes}

A coalescing compact binary in a quasicircular orbit can be completely characterized by its intrinsic
parameters, namely its individual masses $m_i$ and spins $\bm{S}_i$, and its seven extrinsic parameters: right
ascension, declination, luminosity distance, coalescence time, and three Euler angles characterizing its orientation
(e.g., inclination, orbital phase, and polarization).  In this work, we will also use the total mass $M=m_1+m_2$ and mass ratio $q$ defined in the following way:
\begin{equation}
  q = m_2/m_1, \quad \text{where } m_1 \geq m_2.
  \label{eq:q}
\end{equation}
We will also refer to two other commonly used mass parametrizations: the chirp mass $\mc=(m_1 m_2)^{3/5}/(m_1+m_2)^{1/5}$
and the symmetric mass ratio $\eta = m_1 m_2/(m_1+m_2)^2$.
With regard to spin, we define an effective spin \cite{2001PhRvD..64l4013D,2008PhRvD..78d4021R,2011PhRvL.106x1101A},
which is a combination of the spin components along the orbital angular momentum direction $\hat{L}$, in the following way:
\begin{equation}
  \chieff =
  (\bm{S}_{1}/m_{1} + \bm{S}_{2}/m_{2}) \cdot \hat{\bm{L}}/M
  \label{eq:chieff}
\end{equation}
where $\bm{S}_1$ and $\bm{S}_2$ are the spins on the individual BH.  We will also characterize BH spins using the
dimensionless spin variables
\begin{equation}
  \bm{\chi}_{i} = \bm{S}_{i}/m_{i}^{2}.
  \label{eq:chii}
\end{equation}
We will express these dimensionless spins in terms of Cartesian components $\chi_{i,x},\chi_{i,y}, \chi_{i,z}$, expressed
relative to a frame with $\hat{z}=\hat{L}$ and (for simplicity) at the orbital frequency corresponding to the earliest
time of astrophysical interest (e.g., an orbital frequency of $\simeq 10 \, \unit{Hz}$).

When necessary, compact binary parameters are inferred through the use of Bayesian analysis via \textsc{Rapid parameter Inference on gravitational wave sources via Iterative Fitting} (\RIFT{})
\cite{gwastro-PENR-RIFT}, which  reproduces the results of
standard Monte Carlo techniques described in \cite{PEPaper,gw-astro-PE-lalinference-v1} and references therein.
For any  event, fully characterized by parameters $x$, we can compute the (Gaussian) likelihood function $p(d|x)$ for detector network data $d$ containing a signal by using waveform models and an estimate of the (approximately Gaussian) detector noise on short timescales (see, e.g.,~\cite{gw-astro-PE-lalinference-v1,PEPaper,NRPaper} and references therein).
In this expression $x$ is shorthand for the set of 15 parameters needed to fully specify a quasicircular BBH.
The posterior probability distribution is therefore $p(x|d)\propto p(d|x)p(x)$, where $p(x)$ is the prior probability of
finding a merger with different masses, spins, and orientations somewhere in the universe.
These parameters $x$ can and are often described with alternate coordinate systems.  We sometimes refer to
the source luminosity distance $d_L$ or equivalently its source redshift $z$, and to the detector-frame or redshifted
masses $m_{i,z}=m_{i}(1+z)$.  (To distinguish from the detector-frame masses, we will sometimes refer to $m_{i}$ as the
source-frame binary masses.)
LIGO-Virgo analyses have adopted a fiducial prior $p_{\rm ref}(x)$ that is uniform in
orientation, in luminosity distance cubed, in redshifted mass, in spin direction (on the sphere), and, importantly for us, in spin magnitude~\cite{gw-astro-PE-lalinference-v1,PEPaper}.   
Using standard Bayesian tools~\cite{PEPaper,gw-astro-PE-lalinference-v1}, one can produce a sequence of independent, identically distributed samples $x_{n,s}$ ($s=1,2,\ldots,S$) from the posterior distribution $p(x|d)$ for each event $n$;
that is, each $x_{n,s}$ is drawn from a distribution proportional to $p(d_n|x_n)p_{\rm ref}(x_n)$.
Typical calculations of this type provide  $\lesssim 10^4$ samples~\cite{PEPaper,gw-astro-PE-lalinference-v1} from which
the posterior probability distribution is inferred.

For other examples involving purely synthetic observing scenarios, we perform this procedure with  a familiar Fisher
matrix approximation for the form of $p(d|x)$  as a function of $x$ \cite{1994PhRvD..49.2658C,1995PhRvD..52..848P,gwastro-mergers-HeeSuk-FisherMatrixWithAmplitudeCorrections}; see  Appendix \ref{ap:mock} for
details.

\subsection{Population inference}

We use Bayesian inference to constrain the mass and spin distributions of the astrophysical population of BBHs. To do this, we assume that the distribution is one of a family of distributions, parametrized by $\Param$ and scaled by some overall rate $\rate = \mathrm{d}N/(\mathrm{d}t \, \mathrm{d}\Vc)$, which is constant in comoving volume $\Vc$. Each BBH in the population has properties denoted by $\param \equiv (m_1, m_2, \bm{\chi}_1, \bm{\chi}_2)$

Ultimately we are interested in determining the likelihood of the astrophysical BBH population having a given merger rate $\rate$ and obeying a given parametrization $\Param$, given the data for $N$ detections, $\mathcal{D} = (d_1, \ldots, d_N)$. This likelihood, $\mathcal{L}(\rate, \Param) \equiv p(\mathcal{D}\mid\rate, \Param)$, is that of an inhomogeneous Poisson process

\begin{equation}
  \mathcal{L}(\rate, \Param) \propto
  e^{-\mu(\rate, \Param)}
  \prod_{n=1}^N
    \int \mathrm{d}\param \, \ell_n(\param) \, \rate \, p(\param\mid\Param),
  \label{eq:inhomog-poisson-likelihood}
\end{equation}
where $\mu(\rate,\Param)$ is the expected number of detections under a given population parametrization $\Param$ with
overall rate $\rate$ and where $\ell_n(\lambda)=p(d_n|\lambda)$ is the likelihood of data $d_n$ given binary parameters $\lambda$. A derivation for $\mu$ is given in Sec. \ref{sec:sub:VT}.

Using Bayes' theorem, $p(\rate,\Param\mid\mathcal{D}) \propto p(\rate, \Param) \, \mathcal{L}(\rate, \Param)$, one may obtain a posterior distribution on $\rate$ and $\Param$, after assuming some prior $p(\rate, \Param)$. To avoid computing the normalization constant, we instead draw samples from the posterior distribution via Goodman and Weare's affine invariant Markov chain Monte Carlo (MCMC) ensemble sampler \cite{goodman2010}, as implemented in the \textsc{Python} package \textsc{emcee} \cite{2013PASP..125..306F}.

\subsection{Estimate for $VT$}
\label{sec:sub:VT}

Current LIGO-Virgo search sensitivity is well approximated by a familiar approximation: a source will typically be detected if the
estimated signal to noise (SNR) of the second-most-sensitive detector is greater than $8$; see, e.g.,
\cite{LIGO-Inspiral-Rates} and references therein.  Using this
approximation, one can directly evaluate the characteristic volume within which a source will be detected \cite{1993PhRvD..47.2198F}; for nonspinning BH binaries, this estimate is in reasonable agreement with
detailed calculations of search sensitivity   \cite{LIGO-O1-BBH}.
In this work, we therefore adopt the same approximation.
Specifically, we estimate the orientation-averaged sensitive 3-volume $V$ to
which a search is sensitive by the integral \cite{AstroPaper,2010ApJ...716..615O}
\begin{eqnarray}
  V(\lambda) =
  \int
    P(< D(z)/D_h(\lambda))
    \frac{\mathrm{d}V_{\mathrm{c}}}{\mathrm{d}z}
    \frac{\mathrm{d}z}{1+z},
\end{eqnarray}
where $D(z)$ is the luminosity distance for redshift $z$; $D_h(m_1(1+z),m_2(1+z))$ is the horizon distance to which the source can be seen; $V_{\mathrm{c}}$ is the comoving volume;
$z$ is the redshift of the merger event;  and the cumulative distribution $P(>w) = \int_{w>w(\Omega,\iota,\psi)} d\Omega d\psi d\cos \iota$ is a cumulative distribution for
$w=8/\rho$ where $\rho$ is the signal to noise ratio\cite{1993PhRvD..47.2198F,2010ApJ...716..615O,2015ApJ...806..263D}.  Using this definition for $V$, we expect that for a uniform
comoving merger rate $\rate$ (e.g., in units of $\mathrm{Gpc}^{-3} \, \mathrm{yr}^{-1}$), and after observing at this sensitivity for a time $T$ the average number of detections will be
\begin{equation}
  \mu(\rate, \Lambda) =
  \int (VT)(\lambda) \rate p(\param\mid\Param) \mathrm{d}\param,
  \label{eq:mu-from-VT}
\end{equation}
where $p(\param\mid\Param)$ is the probability density function for a random binary in the Universe to have  intrinsic
parameters $\param$.  In this expression,  $\Param$ denotes the parameters that characterize the distribution from which
all coalescing binaries are drawn.
To calculate the horizon distance $D_h$ and hence $V$ for each combination of candidate binary parameters, we use the
\IMRPD  gravitational waveform approximation \cite{2016PhRvD..93d4006H,2016PhRvD..93d4007K}. %

\begin{figure*}
\includegraphics[width=\columnwidth]{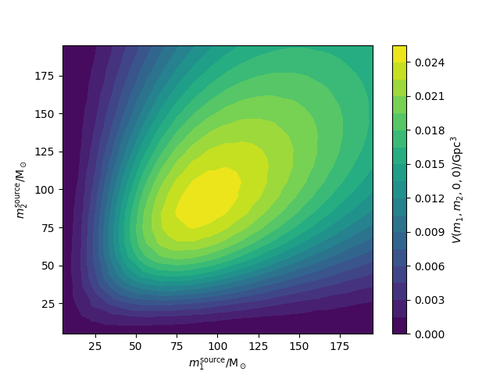}
\includegraphics[width=\columnwidth]{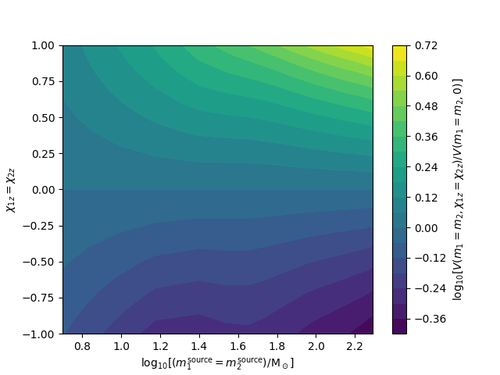}
\caption{Estimated sensitive comoving volume ($V$) versus mass and spin.
\emph{Left:} Sensitive comoving volume $V$ at O1 sensitivity for nonspinning BBHs, in cubic giga-parsecs.
\emph{Right:} Sensitive comoving volume for equal-mass, equal-spin, nonprecessing BBHs, relative to the zero-spin case. Note that $V$ is strictly increased (decreased) if $\chi_{i,z} > 0$ ($< 0$), with higher mass making the effect more pronounced.
\label{fig:V}
}
\end{figure*}

The procedure described above allows us to estimate $V$ for any nonprecessing binary.  Fig. \ref{fig:V} shows this
 estimate as a function of the component masses, based on a single LIGO detector operating at O1 sensitivity.  Motivated by LIGO observations to
date, however, we assume black holes will not be rapidly spinning.  In these circumstances, spin has at best a modest impact
on the sensitive volume; further complications due to precession would be expected to be smaller still
\cite{2012PhRvD..86f4020B,2010PhRvD..82j4006O}.  

Though we pursue a semianalytic estimate for $VT$ and hence the expected number of GW-detected events, detailed analysis of
gravitational wave searches in real data with synthetic sources can evaluate $\mu$ and hence the search sensitivity
directly \cite{2009CQGra..26q5009B,RatesPaper,LIGO-O1-BBH,2018CQGra..35n5009T}.   Such an approach will be particularly necessary when search selection biases (e.g., due to detector noise
non-Gaussianity) cause the search sensitivity threshold to deviate away from the simple SNR threshold described here.

\subsection{Examples of phenomenological population models}

Motivated by the qualitative features of predictions produced by detailed binary formation calculations, several groups
have proposed purely or weakly phenomenological models for  the binary mass distribution
\cite{2017ApJ...851L..25F,LIGO-O1-BBH,2017ApJ...851L..25F,2018ApJ...856..173T,2018ApJ...858L...8C}.   Following \cite{LIGO-O1-BBH,2017ApJ...851L..25F},
we adopt a pure
truncated power law for the relative intrinsic probability $p(m_1,m_2)$ for the source-frame masses in $m_1$ and $m_2$.  
Departing from previous work, we assume the probability density is nonzero only in a region $m_{\mathrm{min}} \leq m_2 \leq m_1 \leq m_{\mathrm{max}}$, and $m_1 + m_2 \leq
M_{\mathrm{max}}$.  Unless otherwise noted,  we assume that $M_{\mathrm{max}}$ is a property of the detector, not
astrophysics, and following the conservative scenario described in \cite{2017ApJ...851L..25F} fix it at $200 \mathrm{M}_\odot$.  
With these assumptions, our mass distribution model has parameters $\alpha_m,k_m,m_{\rm min},m_{\rm max}$ and a functional
form 
\begin{align}
p(m_1,m_2) &= \frac{ (m_2/m_1)^{k_m}  m_1^{-\alpha_m}}{(m_1-m_{\mathrm{min}})} \nonumber \\
 &\times  C(\alpha_m,k_m, m_{\mathrm{min}},m_{\mathrm{max}}, M_{\mathrm{max}})
\end{align}
inside our mass limits and zero elsewhere, 
representing a truncated  power law in $m_1$ with index $-\alpha_m$  and a simple power-law conditional distribution $p(m_2|m_1)$ in secondary mass.   The normalization constant $C$ is defined so $\int_A dm_1 
dm_2 p(m_1,m_2)dm_1 dm_2=1$.  Unless otherwise noted, we will adopt $k_m=0$ in this work.   %
Because GW networks are much more sensitive to more massive BHs with $M\gtrsim 200 \Msun$, this model
and its fiducial choices (e.g., $\alpha_m \simeq 2$) produce a detected merger  distribution $\propto R p(m_1,m_2) VT$ which is roughly uniform
over a wide range of masses, usually terminated by the specific cutoff choices $\mmax,\mmin$ rather than by
selection biases against low mass black holes or the rarity of massive BBHs.   In the  analysis described below, we
leave $\Mmax$ fixed.

Motivated by binary neutron star observations as well as the desire to reproduce arbitrary substructure and features in
the mass distribution, we will also examine Gaussian mass distributions in component mass $m_{i}$
\begin{eqnarray}
p_G(m_1) = \mathcal{N}(\overline{m},\sigma_{m})(m_1)
\end{eqnarray}
which is characterized by its mean value $\overline{m}$ and variance $\sigma_m$.  In this work, we will typically explore the special case of 
$p(m_1,m_2)=p_G(m_1)p_G(m_2)$ and apply this distribution to the case of binary neutron stars, where the narrow
width $\sigma$ relative to the mean $\overline{m}$ implies the distribution has effectively no support for undesirable regions
(e.g., $m<0$). 
Finally, for complete generality, we also discuss  mixtures of mass distributions, including Gaussian mixture
models as previously employed in \cite{WysockiThesis}:
\begin{eqnarray}
p( m_1,m_2|\Lambda) = \sum_\alpha w_\alpha p_\alpha(m_1,m_2|\Lambda_\alpha)
\label{eq:Mixture}
\end{eqnarray}
This latter approach allows complete generality and, with suitable smoothing priors on $w$, the ability to reproduce
arbitrarily complicated mass distributions and circumvent systematic limitations due to our choice of model.  In
particular, these more generic models would allow us to reproduce features previously proposed in
the literature, including overabundances at specific masses near the pair-instability supernova threshold \cite{1968Ap&SS...2...96F,2001ApJ...550..372F,2002RvMP...74.1015W,2007Natur.450..390W,2011ApJ...734..102K,2016A&A...594A..97B}.

For binary black hole spins, we adopt a simple flexible phenomenological model for each BH spin magnitude  $\chi_i$: a  beta distribution, 
\begin{equation}
  p(\chi_i\mid\alpha_{\chi_i}, \beta_{\chi_i}) =
  \frac{
    \chi_i^{\alpha_{\chi_i}-1} \, (\chi_{\rm max}-\chi_i)^{\beta_{\chi_i}-1}
  }{
    \mathrm{B}(\alpha_{\chi_i}, \beta_{\chi_i}) \chi_{\rm max}^{\beta+\alpha+1}
  }
  \label{eq:spin-mag-model}
\end{equation}
with unknown shape parameters $\alpha_{\chi_i}$ and $\beta_{\chi_i}$ ($i=1,2$).   This tractable   two-parameter
distribution allows us to fit to the observed mean and variance---all that the sparse sample of existing
observations will allow.   In this work, we for simplicity assume
both black hole spins are drawn from the same distribution and $\chi_{\rm max}=1$. 
Likewise, for simplicity we adopt the unphysical but easily described 
parametrization of the spin-orbit misalignment $\theta_i = \arccos \hat{\mathbf L}\cdot \hat{\mathbf S}_i$ proposed by
Talbot and Thrane \cite{2017PhRvD..96b3012T}: a unimodal distribution based on a Gaussian in $\cos \theta$ that smoothly deforms
into a uniform distribution in the limit of large $\sigma_{\chi_i}$: 
\begin{align}
  p(\cos\theta_i\mid\sigma_{\chi_i})  \propto
    \mathcal{N}(\cos\theta_i; 1, \sigma_{\chi_i}),
  \label{eq:spin-tilt-model}
\end{align}
When using this model, we assume the polar angles $\phi_i$ of each spin vector relative to the orbital angular momentum direction $\hat{L}$
are uniformly distributed between $0,2\pi$.  In this work, we
assume BH spins are drawn from the same spin misalignment distribution $\sigma_{\chi_1}=\sigma_{\chi_2}$.  In this
approach, as in our parameter inference, all spins are assumed specified at a gravitational wave frequency $f_{\rm ref}=20\,\unit{Hz}$.  No compelling reason exists that astrophysical formation processes should cause binaries of
different masses and spins to be drawn  from a single, universal misalignment distribution at an arbitrary reference
frequency $f_{\rm ref}$; see, e.g., \cite{2018PhRvD..97d3014W,2018PhRvL.120o1101R} for  more detailed models. 
That said, this phenomenological approach is qualitatively consistent with the kinds of misalignments produced by binary
SN natal kicks (e.g., $1-\cos \theta_i \lesssim 0.1 $ for BH natal kicks of order $50\,\unit{km/s}$ \cite{gwastro-ConstrainChannels-BoxingDayKicks-Me2017}),
allowing us a simple way to characterize whether observations support or disfavor plausible amounts of spin-orbit
misalignment.

\subsection{Useful phenomenological parameters}
\label{sec:sub:coordinates}
Observations will  constrain combinations of these phenomenological parameters which reflect clear physical features
in the \emph{observed} (selection-biased) distribution of binary black holes.   We can better characterize what we learn
from GW observations early on by adopting coordinates conforming to these features.  

For example, we could have mixture model [Eq. (\ref{eq:Mixture})] consisting only of elements with distinctive features,
each characterizing a distinctive subpopulation of BHs.  Such subpopulations might be BHs near the pair-instability
supernova peak,  binary neutron stars, and a population of binaries with a continuous mass spectrum formed through hierarchical growth in globular
clusters (see, e.g., \cite{2002MNRAS.330..232C,2017PhRvD..95l4046G,2017ApJ...851L..25F} and references therein).   In such a scenario, observations quickly constrain each element, leveraging their distinctive features to
identify the relative rates ${\cal R}w_\alpha$ and the subpopulations from each domain to constrain that region's
parameters.  For the first few tens of events, these observations will principally constrain the mean and variance of
the \emph{detection-weighted} subpopulation $p_\alpha(m_1,m_2)VT$.    We therefore expect that the following
coordinate system will produce roughly uncorrelated observables, for a typical model: 
(a) the relative rates ${\cal R}w$ for different subpopulations;
(b) the mean chirp mass $\overline{\mc}_\alpha$, symmetric mass ratio $\overline{\eta}_\alpha$,  effective spin
$\overline{\chi}_{\rm eff,\alpha}$, and mean spin $\overline{\chi}$ in each subpopulation, based on our understanding of GW measurement errors; 
and (c) the respective widths $\Sigma_{\mc,\alpha}$, $\Sigma_{\eta,\alpha}$, $\Sigma_{\chi_{\mathrm{eff}},\alpha}$, $\Sigma_\chi$, where
we adopt uppercase to distinguish between these symbols and our model hyperparameters.
In Appendix \ref{ap:UnderstandLimits}, we use order-of-magnitude arguments to explain how reliably each of these
quantities can be measured. %

In the context of our fiducial single-component model, we adopt a reference mass $m_1=m_{\rm ref} = 15 M_\odot$
and
characterize the overall event rate not by its normalization, which depends on unobserved binaries with high and low
masses, but by the event rate ${\cal R} p(m_{\rm ref})$  of binaries whose primary $m_1$ has a mass comparable to
GW151226 \cite{2016PhRvL.116x1103A}.   
We identify other natural coordinates for the   distribution  of $m_1$ via its  detection-weighted cumulative distribution  ${\cal
  P}(<m_1)$:
\begin{eqnarray}
{\cal P}(<x)  = \frac{\int d\lambda V(\lambda) p(\lambda) \Theta(x-m_1(\lambda))}{
\int d\lambda V(\lambda) p(\lambda) 
}
\end{eqnarray}
The mass corresponding to the upper (lower) bound of the 90\% symmetric
detection-weighted probability on $m_1$ serves as a proxy for
$\mmax$ ($\mmin$) which is directly observable and thus a
more natural coordinate.\footnote{By contrast, Talbot
  and Thrane \cite{2018ApJ...856..173T} introduce a model which depends on both a minimum mass $\mmin$ and a tapering mass scale
  $\delta m$, but only a linear combination of them is easily observable; see their Fig. 5.
}  
In this work, we emphasize the upper bound $m_*$ of the detection-weighted mass distribution:
\begin{eqnarray}
{\cal P}(m_*) \equiv 0.95
\end{eqnarray}
For BH spins,   closed-form expressions for
the appropriate mean values and  variances are generally not available for arbitrary selection biases
$VT$; however, to the extent that $VT$ depends only weakly  on BH spin, our model for BH spins and misalignments
[Eqs. (\ref{eq:spin-mag-model},\ref{eq:spin-tilt-model})] implies that
\begin{subequations}
\begin{align}
\overline{\chi}  &\simeq  \frac{\alpha_\chi}{\alpha_\chi+\beta_\chi}\\
\Sigma_{\chi}^2  &\simeq  \frac{\alpha_\chi \beta_\chi}{(\alpha_\chi+\beta_\chi)^2(\alpha_\chi+\beta_\chi+1)}\\ 
\overline{\chi}_{\rm eff} & \simeq  \overline{\chi} \overline{\cos \theta} \\
\overline{\cos \theta} &\simeq \frac{{\rm erf}(\sqrt{2}/\sigma) + 2\sigma(e^{-2/\sigma^2}-1)/\sqrt{2\pi}}{{\rm erf}(\sqrt{2}/\sigma)}
\end{align}
\label{eq:CharacteristicVariables}
\end{subequations}
for our fiducial case where both BH spins are drawn from the same distributions; in these expressions, $\Sigma_\chi^2$
refers to the variance of the one-dimensional $\chi$ distribution, while $\bar{\chi}$ refers to its mean.

\subsection{Interpreting results: Posterior predictive distributions and revised priors}

If we ask any question about compact binary properties $x$ rather than model hyperparameters $\Lambda$, the only quantity that appears in  our
posterior inferences $p(\Lambda|\{d\})$ informed by our observations $\{d\}$  is the posterior
predictive distribution $p_{\rm ppd}(x|\{d\})$:
\begin{eqnarray}
p_{ppd}(x|\{d\}) = %
\int d\Lambda p(x|\Lambda)p(\Lambda|\{d_k\})
\end{eqnarray}
The posterior predictive distribution (PPD) encodes our best estimates of the properties of any randomly selected future
binary, based on observations to date and accounting for our initial prior knowledge about $\Lambda$.  
Unlike the model parameters themselves, which may be highly degenerate and lack physical meaning, the PPD provides an
unambiguous estimate for how likely different binary parameters are, given our knowledge.
Note that by design, the PPD is a probability distribution and,  folding in all uncertainties, does not have an error
estimate.

As events accumulate, we can use  posterior constraints $p(\Lambda| \{d\}_k)$ on model hyperparameters $\Lambda$ based on the first $k=1 \cdots N$
observations to provide a nuanced, observationally revised perspective on future measurements $k>N$.   These prior
insights can be particularly powerful when individual future measurements are only weakly informative about certain
binary parameters such as the mass ratio or spin; see, e.g., \cite{2017PhRvL.119y1103V,2017PhRvD..96l4041W} for  examples.  

To be concrete, our usual population inferences are performed using a single fiducial choice of reference prior $p_{\rm
  ref}(x) = p(x|\Lambda_{\rm ref})$: the posterior is  $p(x|d_k,\Lambda_*) = p(d_k|x)p(x|\Lambda_*) / \int
p(d|x)p(x|\Lambda_{\rm ref})$.   We exploit prior measurements via
\begin{eqnarray}
p(x|d_k,\{d\}) = \frac{p(d_k|x) \int d\Lambda p(x|\Lambda)p(\Lambda|\{d_k\})
}{
\int dx p(d_k|x) \int d\Lambda p(x|\Lambda)p(\Lambda|\{d_k\})
}
\label{eq:PriorInformationUsed}
\end{eqnarray}
In this expression,  the numerator $\int d\Lambda p(x|\Lambda)p(\Lambda|\{d_k\})$ is  the posterior predictive
distribution described above.

\section{Controlled tests with synthetic populations and measurements}
\label{sec:tests1}

To demonstrate our method can infer population parameters, we perform several validation studies using toy models which
mimic key features of real gravitational wave observations.   
These completely controlled illustrations also let us highlight what can be inferred and why about the mass and spin
distribution, within the context of our approach.  
Finally, these examples allow us to demonstrate how population inference can strongly inform the interpretation of
individual future GW observations.

\subsection{BNS mass and (aligned) spin distribution}
\label{sec:sub:bns}
For each component of a binary neutron star (BNS), observations of galactic pulsars suggest that the component masses are
drawn from a Gaussian distribution with mean $1.33 M_\odot$ and standard deviation $0.09 M_\odot$
\cite{2016ARA&A..54..401O}. Observations of pulsars and theoretical models of pulsar spin-down suggest that if both NS are
not recycled, then their dimensionless spins will be small [$\simeq \mathcal{O}(0.05)$]. 
Under the assumption that NS spins are parallel to their orbital angular momentum,  we construct a synthetic population drawn from this phenomenological model; construct synthetic observations for each
binary, recovering 13 synthetic sources based on a three-detector advanced LIGO/Virgo network using a
threshold set by the second-most-sensitive detector's recovered amplitude; perform full GW inference on each source using RIFT \cite{gwastro-PENR-RIFT}; and, with the
resulting posterior distributions, use the techniques of Sec. \ref{sec:method} to infer the underlying NS mass and
spin distribution.  In our reconstruction, we assume both components of a NS binary are independently  drawn from a
Gaussian distribution with unknown mean and variance; and with spins $\chi_{i,z}$ drawn from a beta distribution with
unknown mean and variance, such that $|\chi_{i,z}|\le 0.05$.

\begin{figure}
  \centering
  \includegraphics[width=\columnwidth]{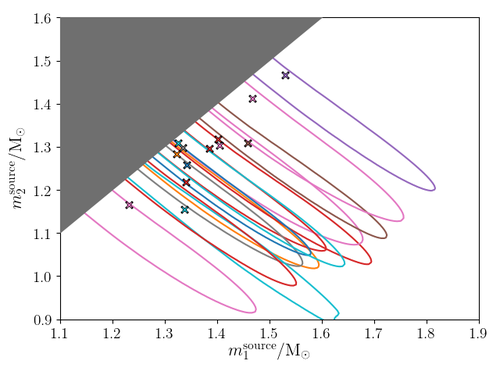}
  \includegraphics[width=\columnwidth]{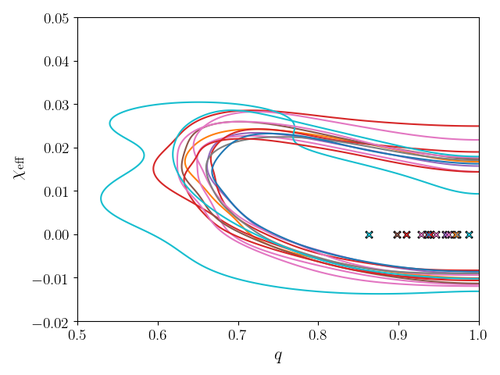}
  \caption{Source information for our synthetic BNS population: For each synthetic signal used in the BNS population
    reconstruction calculation described in Sec. \ref{sec:sub:bns}, these two panels show the true injected source-frame
    parameters (as crosses) and posterior distributions (contours of their 95\% highest posterior density regions).  Each color corresponds to a different source.  Source
    parameters have been inferred using full Bayesian parameter inference via RIFT, as described in the text.
}
  \label{fig:bns-injections}
\end{figure}

\begin{figure}
\includegraphics[width=\columnwidth]{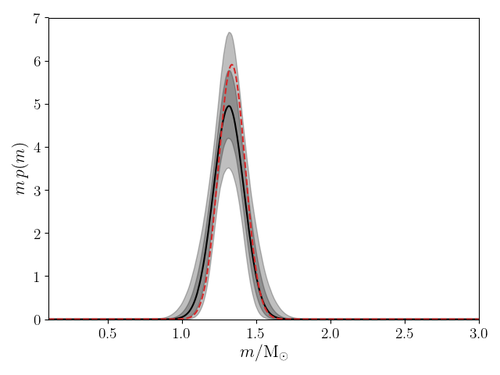}
\includegraphics[width=\columnwidth]{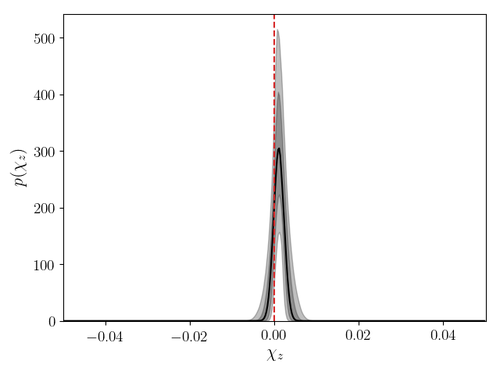}
\caption{Recovered properties of NS mass and spin distribution:  For the synthetic population of BNS sources
  illustrated in Fig. \ref{fig:bns-injections}, this figure shows the recovered mass distribution (top figure) and
  spin distribution parameters (bottom figure) derived using the Gaussian mass and
  $\beta$-distribution spin model described in the text.  The solid line indicates the median distribution; the shaded
  regions indicate the 68\% and 95\% credible intervals.
  Red dashed lines denote the true underlying distribution.  In the case of spin, note that the truth is a delta function at zero, so it would require an infinite number of detections to fall within the constraints on this plot.
}
\label{fig:bns-recover-distribs}
\end{figure}

Fig. \ref{fig:bns-injections}  shows the synthetic measurements used as inputs in our calculation.
 These synthetic measurements incorporate significant uncertainty in each source's redshift, which
  contributes to the overall uncertainty in each binary's chirp mass.
 For each neutron star in our synthetic
population, we use the APR4 equation of state to calculate each neutron star's tidal deformability
$\lambda_i=\lambda(m|\text{APR4})$.   We generate and recover our synthetic sources with
\IMRPD{}\textsc{\_NRTidal} \cite{nrtidal}.
Fig. \ref{fig:bns-recover-distribs} compares our recovered NS mass and spin distribution.  When inferring source parameters, our waveform model and
parameter inferences include the effects of NS tides, treating each NS tidal deformability $\lambda_i$ as a free
parameter.   Despite considerable uncertainties in each
measurement, each BNS observation constrains that binary's chirp  mass reasonably well, to an accuracy $ \sigma_{\mc}  \simeq
0.05 M_\odot$, dominated by uncertainty in source redshift.   Because GW measurements
are only weakly informative about the mass ratio, these measurements each constrain the total mass to be $m_1+m_2 \simeq
2^{6/5} \mc$ to an accuracy $\sigma_{\mc} 2^{6/5}$; averaging all such observations, we can deduce the mean NS mass
$\bar{m}$. With $n=13$ such measurements, we expect to constrain the mean  mass of the population to a
1 standard deviation accuracy
$\sqrt{\sigma_{\mc}^2 2^{12/5}/4 + \sigma^2}/\sqrt{n} \simeq 0.027 M_\odot$, which  compares favorably to $0.02 M_\odot$, the standard
deviation of our Bayesian estimate for $\bar{m}$ .   (A
similar analysis shows that we constrain the NS population standard deviation $\sigma_m$  almost entirely  through
these one-dimensional chirp mass constraints.)   Because GW measurements have a smaller statistical uncertainty than the
astrophysical population width in total mass, the accuracy to which we constrain the mean NS mass is dominated by a simple frequentist
error estimate ($\sigma/\sqrt{n}$), allowing us to reliably project the information we will extract about NS masses from
future GW observations.  

The measurement accuracy for GW measurements of BNS has been long known \cite{1995PhRvD..52..848P}, and their implications for
astrophysics (e.g., mass and BNS spin distributions) have been immediately apparent; see, e.g.,
\cite{gwastro-mergers-HeeSuk-CompareToPE-Aligned,2013ApJ...766L..14H,2018PhRvD..98d3002Z} and references therein.
We provide the first end-to-end demonstration of how well binary NS population parameters can be measured, using a
detailed waveform model at a level where waveform systematics should not dramatically impact the mass, spin, or tidal parameter
inferences being performed.   By contrast, many previous studies focusing on NS tidal deformation have demonstrated that
waveform systematics could bias inferences  \cite{2014PhRvD..89j3012W,2014PhRvL.112j1101F,2015PhRvD..91d3002L}, 
if not controlled.
Only recently have systematic errors between waveform models diminished enough to enable consistent infererence; see,
e.g., \cite{LIGO-GW170817-SourceProperties}.

\begin{figure}
\includegraphics[width=\columnwidth]{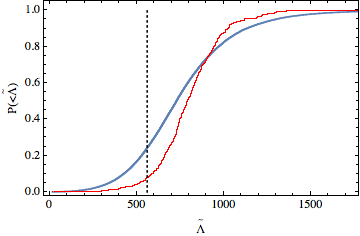}
\caption{Population measurement enables sharper constraints on NS tides:
Cumulative posterior distribution of $\tilde{\Lambda}$ for one of the synthetic sources in our BNS population model.
Blue curve shows a single-event analysis, not exploiting information about the mass and spin distribution from other
events; red curve shows an analysis  based on Eq. (\ref{eq:PriorInformationUsed})  that employs our best estimate for the underlying mass and spin distribution, as
constrained from the population of events in our BNS synthetic sample. 
}
\label{fig:bns:PopulationPriorResult}
\end{figure}

Reliable population inference allows us to draw informed conclusions about future measurements, using previous
observations as prior input.  Particularly for cases like NS binaries where individual measurements can be
weakly informative and produce highly correlated constraints on NS parameters, these prior inputs enable much sharper constraints on astrophysical
parameters.  As a concrete example, Fig. \ref{fig:bns:PopulationPriorResult} shows inferences about one parameter
($\tilde{\Lambda}=
\frac{16}{13}[
(m_1+12 m_2)m_1^4\lambda_1 + (m_2 + 12 m_1) m_2^4\lambda_2]/
(m_1+m_2)^5$) of one of our synthetic NS binaries, where the inferences are performed in isolation (blue line)  and using information obtained from all other
NS observations in our sample about NS masses and spins (but not tides $\tilde{\Lambda}$, which are presumed arbitrary
and spin).  Because our other measurements have allowed us to strongly constrain the NS population's
mass and spin distribution,
we can exploit correlations between our inferences about these parameters and the NS tidal deformability to more
tightly constrain this parameter.  In this way, even though only the strongest few GW measurements will provide most of
the  information about NS tides and the nuclear EOS, by exploiting population measurements we expect to more efficiently draw
conclusions using all available information about the NS population.

\subsection{BBH mass and (precessing) spin distribution}
\label{sec:sub:RecoverBBH}

\begin{table*}
\begin{tabular}{l|cccc|ccc}
Quantity                 & ${\mathcal R}$ & $\alpha_m$ &   $m_{\mathrm{min}}$ & $m_{\mathrm{max}}$ &   $\alpha_\chi$ &   $\beta_\chi$ &$\sigma_\chi$   \\
                       & $\unit{Gpc}^{-3}\unit{yr}$ &  &   $M_\odot$ & $M_\odot$ &   &   &    \\
\hline
Synthetic population &  100 & 0.8 & 5 & 40 & 1.1 & 5.5 & 0.4\\ 
Prior range & $[10^{-1},10^6]$ & $[-5,5]$ & $[5,5]$ & $[30,195]$ & $[10^{-4},10^4]$  & $[10^{-4},10^4]$ & $[10^{-2},10^2]$\\
Prior distribution &  Log-uniform & Uniform & Uniform & Uniform & Log-uniform & Log-uniform & Log-uniform
\end{tabular}
\caption{Synthetic BBH population model: This table shows the parameters of the population model family we adopt to
  generate and recover a synthetic binary black hole population as described in Sec. \ref{sec:sub:RecoverBBH}.  The
  population is characterized by an overall BBH merger rate ${\cal R}$; a power-law slope $\alpha_m$ for the primary
  mass, between $m_{\rm min}$ and $m_{\rm max}$; a beta distribution for spin magnitude, characterized by the two
  parameters $\alpha, \beta$ [Eq (\ref{eq:spin-mag-model})]; and a characteristic misalignment  $\sigma_\chi$ for the angle between BH spins and the
  orbital angular momentum at our reference frequency [Eq. (\ref{eq:spin-tilt-model})].  This analysis also fixes the
  maximum allowed total mass $M_{\rm max}$ (i.e., $m_1+m_2 \le M_{\rm max}$) to $200 M_\odot$.  In this model, both black hole spins are assumed drawn
  independently from the same distribution.  The second row shows the values of these parameters used to
  generate our synthetic population.  The third row shows the range of parameter space we explore when attempting to
  reproduce our data.  The fourth row shows the prior distribution adopted for each parameter, all assumed \emph{a priori}
  independent; in this row, ``log uniform'' implies the prior distribution for any variable $x$ is uniform as a function of $\log x$
  [i.e., $p(x)\propto 1/x$].  Note that for simplicity we have assumed the minimum mass is known.
}
\label{tab:SyntheticBBHParametersAndPriors}

\end{table*}

\begin{figure}
  \centering
  \includegraphics[width=\columnwidth]{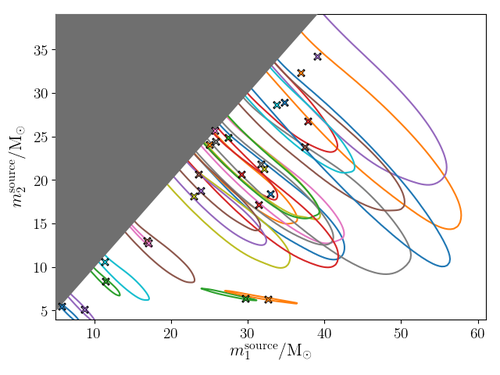}
  \includegraphics[width=\columnwidth]{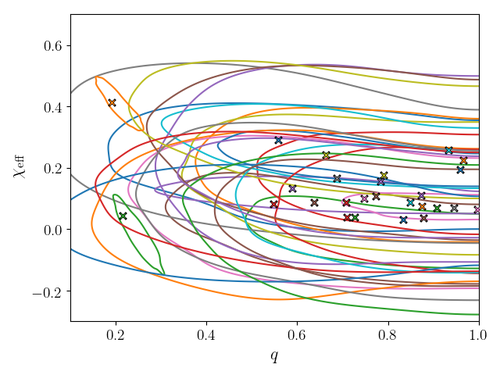}
  \caption{Source information for our synthetic BBH population: For each synthetic signal used in the BH population
    reconstruction calculation described in Sec. \ref{sec:sub:RecoverBBH}, these two panels show the true injected source-frame
    parameters (as crosses) and posterior distributions (contours of their 95\% highest posterior density regions).  Each color corresponds to a different source.
}
  \label{fig:bbh-mock-pe}
\end{figure}

To assess our ability to simultaneously constrain both the mass and spin distribution of binary black holes using GW
observations, we constructed a synthetic population drawn from our fiducial BBH population model, with parameters as
described in Table \ref{tab:SyntheticBBHParametersAndPriors}. 
 Following the procedure described in Appendix
\ref{ap:mockPop}, we drew freely from this population, then selected a subsample based on their relative probability of
detection, producing 25 events based on 300 days of synthetic observation at O1 sensitivity.   For both the synthetic
population and sensitivity model, we approximate $VT$ by neglecting any effects of spin, as a self-consistent
leading-order approximation.    For each event, we generated 1000 fair draws from a synthetic posterior
distribution, using the procedure described in Appendix \ref{ap:mock}.  These synthetic or ``mock'' posterior
distributions mimic the effects of full GW parameter inference, but by construction only explicitly constrain the
binary chirp mass, mass ratio, and effective spin $\chi_{\mathrm{eff}}$ of each event.   Fig. \ref{fig:bbh-mock-pe} shows
the specific source population and synthetic posteriors used in this analysis.   Using these synthetic posterior
distributions, we apply the population inference procedure described in Sec. \ref{sec:method} to produce our best
estimates for the population parameters responsible for our synthetic observations.  
As summarized in Table \ref{tab:SyntheticBBHParametersAndPriors}, our model has parameters
\begin{equation}
\Lambda \equiv (\mathcal{R}, \alpha_m, m_{\mathrm{min}}, m_{\mathrm{max}}, \alpha_\chi, \beta_\chi, \sigma_\chi).
\end{equation}
To be consistent with the priors adopted in other work \cite{LIGO-O1-BBH},  we express our results after reweighting to
correspond to a Jeffries prior on the rate [$\pi(\mathcal{R}) \propto \mathcal{R}^{-1/2}$].
Even with only 25 events drawn from a preferentially low-spin population, our calculations show that GW
measurements should strongly constrain the mass and spin distribution of binary black holes

\begin{figure*}
  \centering
  \includegraphics[width=\columnwidth]{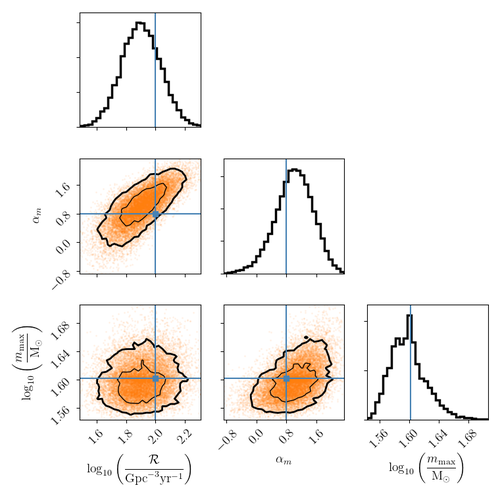}
  \includegraphics[width=\columnwidth]{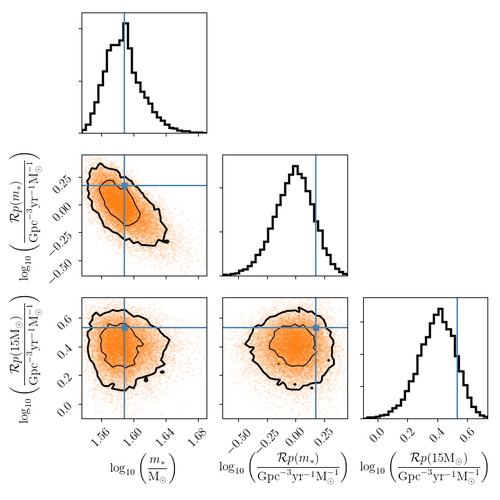}

  \includegraphics[width=\columnwidth]{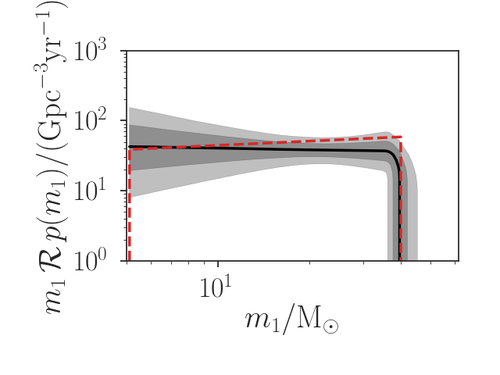}
  \includegraphics[width=\columnwidth]{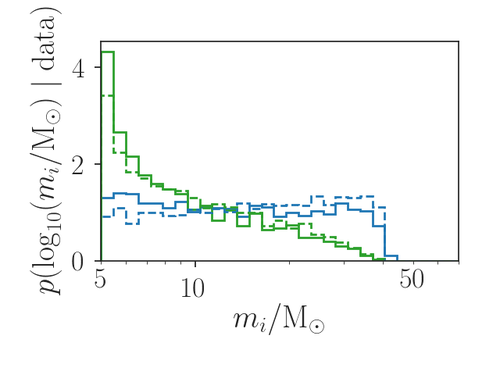}
  \caption{Inferred merger rate versus mass: This figure shows how our estimated merger rate versus mass
    compares with the known distribution used to generate our synthetic source population.  For a more thorough statistical test, see the $P$--$P$ plots in Appendix \ref{ap:pp-plots}.
\emph{Top left}: This group of figures represents the one- and two-dimensional marginal posterior distributions for $\mathcal{R}$, $\alpha_m$, and $m_{\mathrm{max}}$, with the true values overlaid as blue crosshairs.
\emph{Top right}: This group of figures represents the one- and two-dimensional marginal posterior distributions for $m_1^*$, $\mathcal{R} p(m_1^*)$, and $\mathcal{R} p(15\mathrm{M}_\odot)$, with the true values overlaid as blue crosshairs.
\emph{Bottom left}:  In this figure, the red dashed
line shows the characteristic merger rate associated with a given mass scale [$m_1{\cal R} p(m_1)$] versus primary
mass $m_1$.  The black line shows the median inferred value, and the two gray shaded regions show the  symmetric 68\% and 95\% credible
regions.
\emph{Bottom right}: The solid lines in this figure show our posterior predictive distribution $p(m_i| D)$: the best estimate for the probability of a future event being detected having masses $m_i$.  In this figure, blue and green correspond to the primary and secondary masses.  For
comparison, the dotted lines show the true astrophysical distribution.
}
\label{fig:bbh-m1-R-pm1}
\end{figure*}

 Fig. \ref{fig:bbh-m1-R-pm1}  shows how well we can determine the merger rate versus binary masses, such as the primary
 mass.  Notably and in good agreement with previous work, we find we can strongly constrain the maximum detectable
 mass in the population \cite{2017PhRvD..96b3012T,2017ApJ...851L..25F}.  Following the discussion Sec. \ref{sec:sub:coordinates}, however, we emphasize that while the
 maximum \emph{detectable} mass---demarcated by a sharp cutoff in the observed population---is well constrained, the
 \emph{parameters} $\mathcal{R}$, $m_{\mathrm{max}}$, $\alpha_m$ have a degeneracy: as shown in Fig. \ref{fig:bbh-m1-R-pm1}, a population with extremely few but very massive
 BHs is hard to rule out, enabling   larger $m_{\mathrm{max}}$ to be consistent with our synthetic observations.
Additionally and for the first time, we demonstrate how to self-consistently compute both the overall event rate distribution,
including Poisson error, while simultaneously constraining the mass distribution.  Previous investigations have used
specially devised calculations which marginalize over the event rate distribution, producing results that (for a
suitable Jeffries prior) are consistent with our results for the marginal mass distribution.  As desmonstrated in Fig.
\ref{fig:bbh-m1-R-pm1}, to produce a
self-consistent rate distribution, due to strong correlations between the event rate and mass distribution, we must
simultaneously measure the mass-dependent merger rate in the local universe.   Because the correlation between the event rate and mass distribution arises through the
expected number of events, we can provide a simple analytic model for the  correlation between the mass distribution and
event rate, as described  in Appendix
\ref{ap:UnderstandLimits}.

\begin{figure}
  \centering
  \includegraphics[width=\columnwidth]{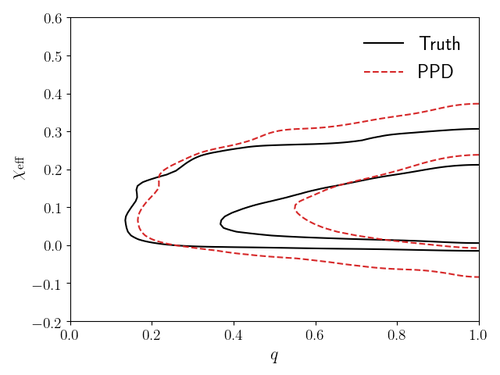}
  \caption{Recovering the true mass ratio and $\chi_{\rm eff}$ distribution: A comparison between the underlying truth (black solid contours) and the inferred posterior predictive (red dashed contours) for the $q$, $\chieff$ marginal distribution.  The inner (outer) contour for each denotes the 50\% (90\%) highest probability density credible region.
}
\label{fig:bbh-ppd-q_chieff}
\end{figure}

With 25 events, our population model has enough information to produce strong constraints on the underlying
phenomenological distributions, even for parameters such as spin which are weakly constrained by individual measurements.
Fig. \ref{fig:bbh-ppd-q_chieff} illustrates how informative these constraints can be about the spin distribution.
This figure compares the true marginal distribution of $q,\chieff$ for the BH-BH population to our best
(posterior predictive) estimate of that distribution.  Even with only a few tens of detections, the estimate traces the general structure of the true distribution. In particular, we can clearly and unambiguously identify that a bias in the  $\chieff$ distribution  toward positive values
suggests an underlying tendency toward alignment.
Of course,  our synthetic observations were intentionally
drawn from the model family we use to fit it; in general, the underlying astrophysical distribution may have a form
outside the model family we adopt, introducing small biases into our interpretation.
Nonetheless, our analysis substantially generalizes previous proof-of-concept demonstrations on how well BH measurements can measure
BH spin distributions, not being limited to a single spin magnitude, a discrete and restrictive family of orientation
distributions, or similar strong prior adopted in previous investigations \cite{2017MNRAS.471.2801S,gwastro-PE-Salvo-EvidenceForAlignment-2015}.

Even with only 25 events, we strongly constrain
the BH spin distribution, in both magnitude and orientation (Fig. \ref{fig:bbh-spin}).  
 As described  in Appendix
\ref{ap:UnderstandLimits} in greater quantitative detail, these two constraints are easily understood.  For this synthetic analysis, the upper limit on spin follows
from the  $\chi_{\rm eff}$ distribution of recovered sources.   Since our synthetic observations included no
events with large $\chi_{\rm eff}$, we can be confident BH spins are not extremely large, since by chance we ought to have found
one large value of $\chi_{\rm eff}$ out of $25$, even allowing for uncertainty in how they are oriented.   
Similarly, because our synthetic population is preferentially aligned ($\sigma=0.4$), the recovered population shown in Fig.
\ref{fig:bns-injections} has a  $\chi_{\rm eff}$ distribution biased toward positive values.  Using
Eq. (\ref{eq:CharacteristicVariables}) for $\overline{\chi}_{\rm eff}$, the bias in  $\chi_{\rm eff}$ inevitably implies 
$\cos \theta$ is preferentially positive and, as described in  Appendix \ref{ap:UnderstandLimits},  allows us to  limit $\sigma$.

\begin{figure}
  \centering
  \includegraphics[width=\columnwidth]{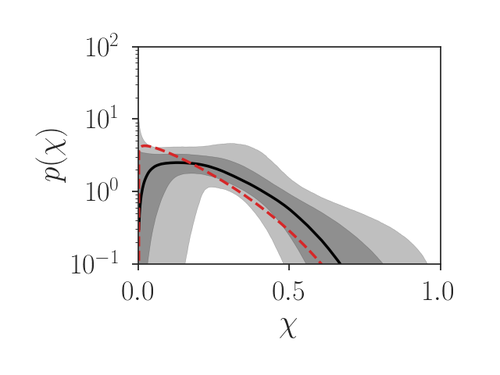}
  \includegraphics[width=\columnwidth]{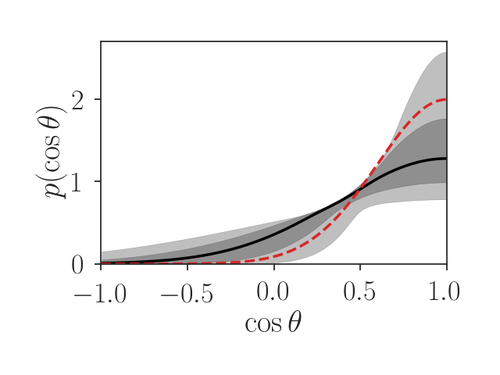}
  \caption{Inferred spin distribution derived from synthetic BBH observations: The top panel shows our
    inferences about the total BH spin; the bottom panel shows our inferences about BH spin-orbit misalignment.
In both panels, the red dashed lines show the underlying distribution, while the black solid lines and shaded regions show the
median recovered parameter distribution.
 To a first approximation, the constraints on spin magnitude and misalignment are as needed
for the population model to reproduce the mass and $\chi_{\rm eff}$ distribution of the underlying population as shown
in Fig. \ref{fig:bbh-ppd-q_chieff}.
}
\label{fig:bbh-spin}
\end{figure}

In this analysis, we employ  conservative synthetic posteriors which  assume only the chirp mass, mass ratio, and effective spin
can be constrained with GW measurements.   
Precessing, coalescing binaries can produce a rich symphony of gravitational waves just prior to and during merger,
reflecting complex binary dynamics and strong-field multimodal radiation.  Given the high expected event rate in ongoing
gravitational wave surveys, we expect that future observations will provide clear examples of precessional dynamics, if
nature produces them, and that these measurements will allow us to much more sharply constrain the BH spin
distribution.  However, for massive BH binaries, model systematics complicate attempts to measure BH parameters,
including spin.  We will conduct full end-to-end calculations with synthetic data and state of the art models in future work.

\section{Analysis of reported observational results}
\label{sec:RealEventAnalysis}

To date, five confident binary black hole mergers have been reported:
GW150914~\cite{2016PhRvL.116f1102A}, GW151226~\cite{2016PhRvL.116x1103A}, GW170104~\cite{2017PhRvL.118v1101A}, GW170608~\cite{LIGO-GW170608}, and GW170814~\cite{LIGO-GW170814} -- 
the latter discovered jointly with the Advanced Virgo instrument~\cite{gw-detectors-Virgo-new},
Additionally, an astrophysically plausible candidate BBH signal   has been reported (LVT151012)~\cite{LIGO-O1-BBH}.
In this section, we describe inferences about the binary black hole population based on reported events, deduced from
these reported observations and a simplified model for the network's search sensitivity.  For
O1 events, most notably for GW151226, we use full posterior inferences derived from GW data, provided by the
  LIGO Scientific Collaboration.
For O2 events, in lieu of full posterior inferences, we use the procedure described in Appendix \ref{ap:mock} to generate synthetic posterior distributions which closely
resemble the reported parameter estimates for mass and $\chi_{\rm eff}$.
For simplicity as well as to enable a concrete illustration of our method using real data, we will produce estimates under the (unwarranted) assumption that
reported O2 results available to date represent a comprehensive and   fair sample of binary black holes seen during
LIGO's O2 observing run.  In these estimates, we assume O1 and O2 share a common sensitive volume $V$ as estimated in
Sec. \ref{sec:sub:VT}, with observing duration
$T_{\mathrm{O1}}=48.6\,\unit{days}$  \cite{LIGO-O1-BBH} and
$T_{\mathrm{O2}}=117\,\unit{days}$  \cite{LIGO-GW170817-bns}.
Keeping in mind model systematics such as the omission of a salient feature in the mass distribution can demonstrably strongly bias
recovered model parameters \cite{2017ApJ...851L..25F,2018ApJ...856..173T}, as well as sample incompleteness for our
O2-scale analysis, in Table \ref{tab:O1O2:ParameterConstraints}
we provide our inferences about the O1 and O2 population within the context of the fiducial BBH population model
described in Sec. \ref{sec:sub:RecoverBBH}.  
For O2 in particular, we emphasize the simplified $VT$ and non final sample used in that analysis, which is provided
solely for illustration and to connect to previously published investigations about O2-scale events \cite{2017ApJ...851L..25F,2017Natur.548..426F,2018PhRvD..97d3014W};  applying our methods to final O2 results with real samples and carefully calibrated $VT$
could produce  substantially different astrophysical conclusions.

\begin{figure*}
\centering
\includegraphics[width=\columnwidth]{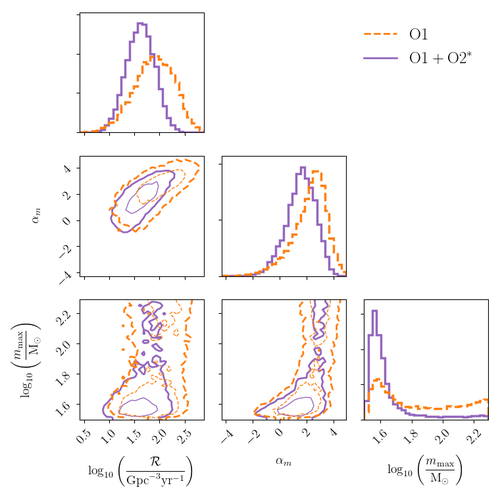}
\includegraphics[width=\columnwidth]{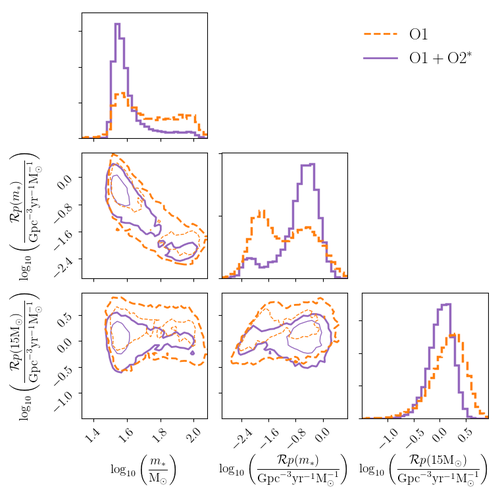}
\caption{Inferences about astrophysical binary BH mass distribution: Inferences about
  the merger rate versus mass  of coalescing BH-BH binaries, using only
  O1 observations (dashed orange) and using O1 and reported O2 observations (solid purple), for simplicity assuming the latter represent a comprehensive and
  fair sample.  We apply an asterisk (O2\textsuperscript{*}) to all O2 results, to highlight the nonfinal sample,
  simplified sensitivity model $VT$, and mocked-up posteriors used in this
  proof-of-concept analysis.  The panels in this figure follow the format of Fig. \ref{fig:bbh-m1-R-pm1} for
  representing one- and two-dimensional marginal posterior distributions.
}
\label{fig:O1:bbh-mass}
\end{figure*}

\begin{figure*}
\includegraphics[width=\columnwidth]{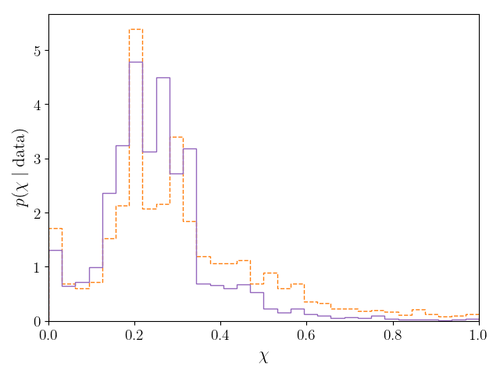}
\includegraphics[width=\columnwidth]{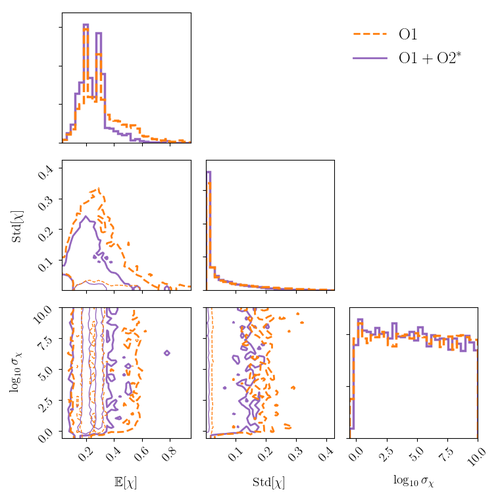}
\caption{Inferences about astrophysical binary BH spin distribution. \emph{Left}: Our best estimates for the binary BH spin
 magnitude distribution (PPD) based on O1 (dashed orange) and O2 (solid purple) observations.  We apply an asterisk
 (O2\textsuperscript{*}) to all O2 results, to highlight the 
non final sample,
  simplified sensitivity model $VT$, and mocked-up posteriors used in this
  proof-of-concept analysis.  Due to the low characteristic spin and
 within the context of the information used in this analysis, these  observations remain uninformative about BH
 spin-orbit orientations.
\emph{Right}: Our best estimates for the binary BH spin distribution, as expressed using our model
hyperparameters, for O1 and O2.
}
\end{figure*}

\begin{table*}
\begin{tabular}{l|cccccccc}
& $\mathcal{R}$ & $\alpha_m$ & $m_{\mathrm{max}}$ & $\mathbb{E}[\chi]$ & $\mathrm{Std}[\chi]$ & $\log_{10}\sigma_\chi$ & $\chi_{\mathrm{eff}}$ & $\chi$ \\
& $\mathrm{Gpc}^{-3} \, \mathrm{yr}^{-1}$ &  & $\mathrm{M}_\odot$ &  &  &  &  &  \\ \hline
O1 & $122^{+291}_{-96}$ & $2.8^{+1.4}_{-2.5}$ & $70^{+110}_{-30}$ & $0.28^{+0.31}_{-0.15}$ & $0.02^{+0.25}_{-0.02}$ & $0.1$--$9.5$ & $0.00^{+0.24}_{-0.24}$ & $0.03$--$0.68$ \\
O2\textsuperscript{*} & $\cdots$ & $1.9^{+1.5}_{-2.0}$ & $39^{+98}_{-6}$ & $0.24^{+0.21}_{-0.12}$ & $0.01^{+0.19}_{-0.01}$ & $0.3$--$9.4$ & $0.00^{+0.19}_{-0.19}$ & $0.04$--$0.49$ \\
\end{tabular}

\caption{Inferences about astrophysical binary BH model parameters: This
  table provides 90\% credible intervals for the  underlying parameters of our fiducial BBH population model, applied to
  O1 and reported O2 observations as described in the text.  Parameters with clear unimodal structure are represented by
  their median and the widths of their 90\% symmetric probability confidence interval, whereas we only report the 90\% upper and lower limits for
  more poorly constrained parameters.  For the spin magnitude distribution, rather than show the (highly correlated)
  credible intervals for the underlying sampling variables $\alpha_\chi,\beta_\chi$, we instead show credible intervals
  for the mean value of $\chi$ and the standard deviation of $\chi$.  We also show the posterior predictive range of
  spin magnitudes $\chi$ and effective spins $\chi_{\mathrm{eff}}$. We apply an asterisk (O2\textsuperscript{*}) to all
  O2 results,
to highlight the nonfinal sample,
  simplified sensitivity model $VT$, and mocked-up posteriors used in this
  proof-of-concept analysis.
}
\label{tab:O1O2:ParameterConstraints}
\end{table*}

Fig. \ref{fig:O1:bbh-mass} shows our best estimates for the merger rate of BH-BH binaries of different masses, inferred
within the context of the model described in Table \ref{tab:SyntheticBBHParametersAndPriors} and demonstrated on
synthetic data in Sec. \ref{sec:sub:RecoverBBH}.    
Naturally, we estimate an overall BH-BH merger rate and mass distribution consistent with previously reported results
\cite{LIGO-O1-BBH}.  Using a Jeffries' prior for the merger rate, we find $\mathcal{R} = 122^{+291}_{-96}\,\unit{Gpc}^{-3}\unit{yr}^{-1}$ based on O1.
For O2, we find uncertainty in the event rate is reduced by roughly a factor of 2, both through reduced Poisson error
(e.g., six instead of three events) and through sharper constraints on the mass distribution (e.g., reducing prospects
for a large maximum mass).  Our result for O1 is
more conservative (wider) than the power-law result reported previously in Abbott \emph{et al.} \cite{LIGO-O1-BBH}, $97^{+135}_{-67}\,\unit{Gpc}^{-3}\unit{yr}^{-1}$,  because we employ a more flexible model and
therefore incorporate more model systematics, notably including the correlation between event rate and mass spectrum and
also the impact of the upper mass cutoff.   Conversely, if we employ consistent assumptions, we arrive at the same
answers previously reported for O1 \cite{LIGO-O1-BBH}.
As we adopt a merger rate model that reduces to previously investigated power laws, by design we reproduce the analysis reported in \cite{2017ApJ...851L..25F}: the events reported during O2 suggest the
absence of very massive BHs in the observable population.\footnote{While our assumptions about the mass distribution
  model have modestly changed relative to Fishbach \emph{et al.} \cite{2017ApJ...851L..25F}, we reproduce their results when adopting the same inputs and
  mass model.}
For this reason our inferences about the mass spectrum exponent $\alpha_m$ are considerably wider than prior work
which does not take a possible upper mass cutoff into account.
Even with the small sample publicly reported so far, our analysis corroborates the analysis in \cite{2017ApJ...851L..25F} that O2-scale GW measurements could be weakly informative about the maximum mass of coalescing BHs.

As demonstrated in several previous investigations \cite{2017Natur.548..426F,2018PhRvD..97d3014W}, we know that BHs in merging binaries likely
have low typical spin.  For example, based on the distribution of $\chi_{\rm eff}$, Farr \emph{et al.} \cite{2017Natur.548..426F} argued that several
members of  a discrete array of
candidate spin  orientations (aligned or isotropic) and magnitude distributions are  inconsistent with observations to date, and that BH spins were likely randomly oriented or small.
Later,  Wysocki and collaborators \cite{2018PhRvD..97d3014W} demonstrated that, if binary black holes arose from
isolated binaries whose spins were weakly misaligned by SN natal kicks, then only relatively small BH natal spins were
consistent with observations available at the time.  
As shown in Fig. 10, with more events available to our analysis, and using much more flexible models, we can draw sharper and more generic
conclusions about the BH spin distribution, even using only six reported events.   
First and foremost, exactly as seen with synthetic data, the absence of large $\chi_{\rm eff}$ allows us to with
increasing confidence bound above the fraction of BHs in merging binaries that have large spin.
Too, because collectively the observed population distribution
of $\chi_{\rm eff}$ remains nearly symmetrically distributed around zero, we can with increasing confidence bound the
fraction of binaries that are preferentially aligned and with modest spin.
With at least one BH known to have spin (GW151226) and for simplicitly assuming the BH spin and mass distribution are
uncorrelated, we are led to weakly disfavor scenarios where BHs are preferentially aligned (i.e., small $\sigma$ is
disfavored).  We emphasize, however, that this conclusion is driven by the absence of strong support for \emph{any} spin
in all but one binary (GW151226).  We would arrive at the same nominal conclusion for a comparable number of random
draws from a binary population model with perfectly aligned binaries with small BH spins.   Future and more informative observations of BH binaries could significantly alter this
conclusion.

\section{Discussion}
\label{sec:discussion}

In this work, we present concrete examples for how well just a handful of GW measurements can improve our phenomenology
of the BH mass and spin distribution.   Our examples include real observational data from LIGO's O1 and (an incomplete sample
from) O2 observing run, suggesting current observations could be on the cusp of constraining BH spins and maximum
masses.  We provide simple estimates to understand how well these parameters have been
constrained, allowing the reader to extrapolate to larger sample sizes.  For example, in the absence of positive support
for spin, the upper limit on BH spin will decrease rapidly, allowing us to place strong upper limits for (or enable
discovery of) BH natal spin.

Because each empirical marginal distribution possesses an infinite number of degrees of freedom, any
phenomenological parametrization such as our own can  quickly be exhausted by the data \cite{2013PhRvD..88h4061O}, particularly when
the population must reproduce multiple observational features.  In the short run, therefore, we anticipate a fully
generic and regularized infinite-dimensional approach will soon be required to adequately reproduce the thousands of events that even
the current generation of instruments will discover. 
A fully generic approach, however, can easily be misled, not least  because  GW measurements are subject to many subtle strong-field systematics due to model incompleteness.  For example,
a waveform approximation widely  used for rapid parameter inference of binary black holes (IMRPv2 \cite{2014PhRvL.113o1101H}) omits astrophysically
critical degrees of freedom---the calculation allows for only one precessing spin instead of the two necessary to
fully describe the dynamics---and demonstrably has systematic errors large enough to shift
 posterior distributions for O3-scale events by an appreciable fraction of their statistically expected extent
\cite{2017PhRvD..96l4041W,gwastro-PENR-RIFT}.
To illustrate the pernicious impact of these systematic biases, we can consider a simple order-of-magnitude estimate: a
single quantity, with intrinsic Gaussian distribution of mean $\mu$ and width $\sigma$, being observed multiple times by an apparatus
with a (Gaussian, random) measurement error $\Delta x$ and bias $\delta x$.  The bias will be important when it
influences our best estimate of the average (i.e., when $\delta x
\gtrsim \sqrt{\sigma^2+\Delta x^2}/\sqrt{N}$).  
Applying this order-of-magnitude approach to GW measurements, we expect that
after only a few tens of binary mergers, these modeling systematics will progressively
 contaminate the interpretation of coalescing binaries, as posterior biases in each event become reflected in biases in
 the inferred population distribution.  
Waveform systematics will be even more important because BH spins appear to be small: greater accuracy is
needed to separate the secular effects of spin.  
In this work, when carrying out a full parameter inference, we use  the newly developed \RIFT{} parameter inference engine \cite{gwastro-PENR-RIFT} to produce posteriors.  We will discuss the impact of waveform systematics on BH spin misalignment measurements in future work.

\section{Conclusions}
\label{sec:conclusions}

We have introduced a flexible, ready-to-use, and self-consistent parametric method to estimate the compact binary merger rate as a function
of binary parameters, specifically emphasizing mass and spin.
Unlike prior work, our procedure self-consistently estimates the merger rate and binary parameter distribution,
accounting for statistical sampling error, measurement error, and selection bias.
Using this procedure, we show by example that only a handful of NS-NS and BH-BH measurements can enable strong
constraints on their respective populations via GW observations alone.    Even in the astrophysically likely scenario of
small BH spin, we emphasize that just a few measurements will enable sharp constraints on the BH spin
distribution.
Interpreting current observations,  we show that GW measurements are already beginning to place astrophysically
interesting constraints on the spin of BHs.  We reproduce prior results about the lack of reported BHs at high
mass and its implications for the BH mass spectrum.  
Finally, particularly in our appendix, we explain how to extrapolate toward the measurement prospects available in the
very near future.

The procedure described here assumes all sources have
been unambiguously resolved from observational data, omitting any treatment of source significance aside from a naive
selection bias.  Farr \emph{et al.} \cite{2015PhRvD..91b3005F} demonstrated and popularized
 an approach to  self-consistently perform the detection and population inference process, estimating the foreground and background
distributions simultaneously; see also \cite{2004AIPC..735..195L,2015ApJ...802...89B,2013NJPh...15e3027M}.   Recently,
Gaebel and collaborators \cite{gwastro-FGMCHierarchical-Sebastian-2018} developed a concrete procedure to
apply this technique to gravitational wave observations.  Owing to many deep similarities between our strategies, we
anticipate we will shortly incorporate this technique in our own analysis.

The approach described here also employs several strong assumptions about the (lack of) correlations between model
parameters.  For example, our fiducial BH  model assumes the mass-dependent BH merger rate is independent of redshift;
that BH masses and spins are completely independent; and that BH spin misalignment and spin magnitudes are likewise
uncorrelated.   We will explore more physically motivated correlations in future work.

In the long run, phenomenology is only as sound as the underlying parametrization.  Previous analyses have repeatedly
shown that adopting an overly restrictive model will produce biased results, as demonstrated by Fishbach et al (with the
maximum mass) \cite{2017ApJ...851L..25F} and Talbot  et al \cite{2018ApJ...856..173T} (with the shape of the maximum mass cutoff).  With sufficient data, a suitably regularized
infinite-dimensional parametrization  will make unintended systematic biases less frequent.  Mature methods for
infinite-dimensional or nonparametric inference exist \cite{book-Gelman-BDA,OrbanzTeh2010,book-GhosalVort-NonparametricBayes}, beginning with simple infinite-dimensional
parametrizations plus smoothing priors or with Gaussian processes \cite{book-Rasmussen-GP}.   Early investigations have
applied nonparametric methods to GW population estimates  \cite{WysockiThesis,2017MNRAS.465.3254M}. However, because the GW
signal is so rich,   many parameters can be measured for each event, several of which are believed to be correlated in
most astrophysical formation scenarios.  These correlations should be more sharply identified with strong theoretical
priors for the immediate future.

Finally, several technical improvements can make this approach faster and more robust.  For example, we can perform inference on all events simultaneously, using direct estimates of the likelihood
$\ell(\lambda)$ naturally reported by \RIFT{}, to ensure any population inferences are not limited by the compact support
of fiducial priors.
Using accelerated general-purpose inference engines, we expect to dramatically accelerate the
speed with which our population inferences are provided, with a long-term goal of enabling low-latency
population-informed identification and classification of candidate sources.

\begin{acknowledgments}
The authors appreciate the opportunities to talk about this work during its development with  Maya Fishbach, Tom Dent, Jonah Kanner, Will Farr, Colm Talbot,  Eric
Thrane, and Salvatore Vitale.  R.O.S.,  J.L., and D.W. gratefully acknowledge NSF award PHY-1707965.
D. W. also acknowledges support from the Rochester Institute of Technology through the Frontiers in Gravitational Wave Astrophysics (FGWA) Signature Interdisciplinary Research Areas (SIRA) initiative.
The authors thank the LIGO Scientific Collaboration for access to the data and gratefully acknowledge the support of the United States National Science Foundation (NSF) for the construction and operation of the LIGO Laboratory and Advanced LIGO as well as the Science and Technology Facilities Council (STFC) of the United Kingdom, and the Max-Planck-Society (MPS) for support of the construction of Advanced LIGO. Additional support for Advanced LIGO was provided by the Australian Research Council.
For their use in building the PopModels package, we would like to acknowledge \textsc{Numpy} and \textsc{Scipy} \cite{scipy}, \textsc{Emcee} \cite{2013PASP..125..306F}, \textsc{Matplotlib} \cite{matplotlib}, \textsc{AstroPy} \cite{astropy:2013,astropy:2018}, and \textsc{h5py} \cite{collete_2013}.

\end{acknowledgments}

\appendix

\section{Mock posterior populations precessing binaries: Aligned Fisher matrix approach}
\label{ap:mock}
We test our code using synthetic or ``mock'' posterior distributions for binary black hole parameters, designed to mimic the results
of  full end-to-end Bayesian inference on synthetic data.  For the mock BBH posterior distributions constructed in this
work, we adopt a very simple approximation, motivated by decades of experience suggesting that for short BBH signals the
likelihood for gravitational wave signals is nearly Gaussian in three coordinates ($\mc,\eta,\chi_{\rm eff}$) and does
not strongly constrain any other degrees of freedom.  Specifically, if $\lambda_0$ are the true binary parameters and
$\rho$ is the true network signal amplitude; if
$\Gamma_{ab} = \qmstateproduct{\partial_a h}{\partial_b h}$ is the Fisher matrix for the binary parameters $\lambda$,
evaluated at $\lambda=\lambda_0$ and for a signal amplitude $\rho$ using a fiducial detector power spetcrum; and if
$p(\lambda)$ is the prior distribution on $\lambda$, then we approximate the posterior distribution by a distribution
proportional to 
\begin{eqnarray}
e^{- \Gamma_{ab}(\lambda-\lambda_*)_a(\lambda-\lambda_*)_b}p_{\rm ref}(\lambda)
\end{eqnarray}
where $\lambda_*$ is a fixed random realization from a normal distribution with mean $\lambda_0$ and covariance matrix
$\Gamma^{-1}$.  
We generate samples from this distribution via Monte Carlo techniques.  We evaluate the approximate Fisher matrix $\Gamma$ using the
effective Fisher technique
\cite{gwastro-mergers-HeeSuk-CompareToPE-Aligned,gwastro-mergers-HeeSuk-FisherMatrixWithAmplitudeCorrections,2014CQGra..31w5009C},
applied to a nonprecessing binary waveform model assigned the same values of $\mc,\eta,\chi_{\rm eff}$ (i.e., via
$\chi_{1,z}=\chi_{2,z}=\chi_{\rm eff}$).

This approximate posterior distribution has several distinct advantages.  First and foremost, it captures in
$\Gamma_{ab}$ the strong, 
parameter-dependent, and well-understood correlations between the variables that most significantly impact the GW
inspiral signal, while simultaneously populating all intrinsic binary parameters.  
For example, it captures the shape of the posterior distribution in mass ratio and spin while correctly accounting for
parameter boundary effects, as described in \cite{2018PhRvD..98h3007N}.  
Second, it accounts via $\lambda_*$ for the effect of random noise realizations, which impact the
best-fitting parameters associated with each set of synthetic data.  By including an explicit prior
$p_{\rm ref} (\lambda)$, it allows us to carefully adopt fiducial prior assumptions, which have a substantial impact on
inferred binary masses and spins.

A ready-to-use  implementation of this algorithm is available.\footnote{See
  \href{https://git.ligo.org/daniel.wysocki/synthetic-PE-posteriors}{https://git.ligo.org/daniel.wysocki/synthetic-PE-posteriors}.}

For simplicity, in this implementation, no cosmological effects are applied.  If used unaltered, this approximate
posterior applies either if cosmological redshift effects are small compared to the width of the distribution in mass
(i.e., bias is small compared to the statistical uncertainty) or if these ambiguity distributions are used to approximate
the source-frame ambiguity function.  Cosmological effects dominate the accuracy to which a binary neutron star's chirp
mass can be measured; to be used in such a scenario, this approximation must be refined to reflect the significant
impact of the sources' unknown redshift.

\section{Mock populations}
\label{ap:mockPop}

To generate a synthetic population of events, we employ the following procedure.   
Using O1 sensitivity, and a detection criterion of $\rho > 8$ in a single interferometer, we used our estimate of $V$ and a
fiducial observation time $T$ to compute the expected number of events $\mu$.   Using the Poisson distribution, we
select a total number of events $N$ to observe.  
We assumed each detected binary had a network SNR drawn from a power law $p(\rho_{\mathrm{network}}) \propto \rho_{\mathrm{network}}^{-4}$, with a lower cutoff of $12$ (roughly corresponds to $8$ in two detectors).

\section{Overview of key phenomenological constraints}
\label{ap:UnderstandLimits}

\subsection{How well can we measure distribution hyperparameters?}

Classical frequentist statistical methods provide a quick way to assess how rapidly observations will constrain model
hyperparameters.  For example,  the sample mean of  maximum likelihood estimators converges rapidly to the true
mean, and (to a first approximation) the sample variance is approximately $\chi^2$ distributed.  Thus, by adopting  the mean
and variance of our underlying distributions as coordinates on the space $\Lambda$ of hyperparameters, we can estimate
how efficiently observations will constrain them.   
For example, if we  account for measurement error, we can measure the mean spin to an accuracy
$\sqrt{V(\chi) + \sigma_\chi^2}/\sqrt{N}$ where $V(\chi)$ is the variance of the spin magnitude distribution and
$\sigma_\chi$ is the typical spin measurement accuracy for the mass range of interest [typically $\mathcal{O}(0.3)$].
Because of sharp cutoffs, the maximum and minimum masses have a qualitatively different behavior; see, e.g., \cite{amari2007methods}.  Both the maximum and
minimum masses are best estimated using the most extreme individual event, with an accuracy converging as $1/N$.  In our
context---the power-law mass distribution---the accuracy with which these maximum masses can be determined scales
directly with the number of events in a given region.  We therefore expect the maximum mass can be
determined to an accuracy of order $m_{\rm max}/N$; the appropriate scale factor can be calibrated to detailed analyses of
the kind performed in Sec. \ref{sec:tests1}. 
Similarly, as described below in Appendix \ref{ap:sub:SpinConstraints},  we can use the observed range of $\chi_{\rm
  eff}$ to constrain spin magnitudes and misalignments.

While providing a useful order-of-magnitude estimate into how well we can measure distribution parameters, the simple
estimates above become cumbersome when trying to capture correlations between our phenomenological parameters, notably
the event rate and mass distribution.  
Following  \cite{2013PhRvD..88h4061O}, we assess how well we can distinguish model hyperparameters 
from the (expected) log-likelihood as a function of model hyperparameters $\Lambda$
of 
\begin{eqnarray}
\E{\ln {\cal L}} =  -   \mu_*  + \mu_* \E{\ln \int d\lambda p(d|\lambda) R p(\lambda|\Lambda)}_* 
\end{eqnarray}
where the expectation is performed relative to some reference model characterized by parameters $\Lambda_*$ such that
$p_*(\lambda)\equiv p(\lambda|\Lambda_*)$ and $\mu_*=\mu(\Lambda_*)$.  
Rather than work in full generality, we perform a Taylor series expansion of the likelihood around the local maximum,
characterizing the second order term by its inverse covariance or Fisher matrix $\Gamma_{ab}$
\begin{eqnarray}
\E{\ln {\cal L}} \simeq \ln {\cal L}_*  - \frac{1}{2}\Gamma_{\alpha \beta} (\Lambda-\Lambda_*)_\alpha (\Lambda-\Lambda_*)_\beta
\end{eqnarray}
If $\gamma_k$ are eigenvalues of $\Gamma$, then hyperparameters can be measured to an accuracy $1/\sqrt{\gamma_k}$,
which scales as $1/\sqrt{N}$ for $N$ the number of observed events.

We first illustrate this technique in the idealized case of zero
measurement error, following  previous work \cite{2013PhRvD..88h4061O} which  characterized 
 differences between
two distributions $q,p$ using
the KL divergence $D_{\mathrm{KL}}(p|q) \equiv \int p(x) \ln [p(x)/q(x)] \mathrm{d}x$.  
The marginalized log likelihood only depends on
model hyperparameters $\Lambda$ through  the KL divergence between our proposed model $\mu,p$ (which depends on $\Lambda$)
and the  reference model $\mu_*,p_*$ (which does not):
\begin{eqnarray}
\E{\ln {\cal L}} = - D_{\mathrm{KL}}(\mu_*|\mu) - \mu_* D_{\mathrm{KL}}(p_*|p) + \text{const.}
\end{eqnarray}
As a result, the Fisher matrix has two model-dependent terms, each reflecting second derivatives of $D_{\mathrm{KL}}$ with
respect to model parameters:
\begin{eqnarray}
\Gamma_{\alpha\beta}^{(zero)} = \Gamma_{\alpha,\beta}^{(\mu)} + \mu_* \Gamma_{\alpha \beta}^{(p)}
\end{eqnarray}
where the first term arises from differences in the observed number; where the second term reflects differences in shape; and
where we use the fact that $D_{\mathrm{KL}}$ has a local minimum (of 0) when the two distributions are equal to eliminate cross
terms.  
Thus, we can evaluate the Fisher matrix simply by computing KL divergences and carrying out the
necessary derivatives.  For example, for the mass power-law model with fixed mass range, $p(m|\alpha) =C(\alpha,m_+,m_-)
m_1^{-\alpha}/(m_1-m_-)$, the KL divergence $D_{\mathrm{KL}}(p_*,p)$ becomes
\begin{align}
D_{\mathrm{KL}}(\alpha_*|\alpha) 
 & \equiv \int p(x|\alpha_*) \ln p(x|\alpha_*)/p(x|\alpha) \\
 &=(\alpha - \alpha_*) \E{\ln x}_{\alpha_*} +  \ln C(\alpha_*)/C(\alpha)
\end{align}
where the conditional average is $\E{f}_\alpha \equiv \int dx f(x)p(x|\alpha)$.  
In this expression, only the last term $-\ln C(\alpha)$ does not cancel in $\partial_\alpha^2 D_{\mathrm{KL}}(\alpha_*|\alpha)$.  

Again using the same concrete power-law example,   we next use this technique to show how, because $\mu$
[Eq. (\ref{eq:mu-from-VT})] and the mass
distribution can be independently constrained, the ``overall event rate'' ${\cal R}$ and the mass distribution are
correlated.  Representing $\mu = e^X$, the second derivative of $D_{\mathrm{KL}}(\mu_*|\mu)$ becomes \cite{2013PhRvD..88h4061O}
\begin{eqnarray}
D_{KL} \simeq \frac{1}{2}  \mu_*   (\partial_a X)(\partial_b X)(\Lambda-\Lambda_*)_a(\Lambda-\Lambda_*)_b
\end{eqnarray}
For the power-law model described above, the only two derivatives needed are $\partial_{\ln R} X =1$ and
$\partial_{\alpha} X = \partial_\alpha \ln \E{VT}_{\alpha}$, the latter of which can be well approximated by $-1$.  
This term introduces correlations between the rate variable ($\ln {\cal R}$) and shape ($\alpha$).  Conversely, using
coordinates $\mu$ and $\alpha$ to characterize the observed population, by construction our inferred posterior distribution on
the total number and mass distribution are uncorrelated. 

Roughly speaking, the effects of measurement error add in quadrature in the Fisher matrix:
\begin{eqnarray}
\Gamma =\Gamma^{(\mathrm{zero})} + \Gamma^{(\mathrm{measure})}
\end{eqnarray}
We can therefore refine the estimates provided above to incorporate simple estimates of GW measurement errors and their
correlations.  For the simple power-law estimate described above, however, these measurement errors are relatively small
compared to the range of the distribution, unless $\alpha$ is very large.  

In the above order-of-magnitude discussion, we have not accounted for parameter-dependent selection bias.  To a good first approximation,
GW selection bias enters only through the masses, roughly as the (chirp) mass to a power.  We can therefore treat the
observed population as a (different) power law, which observations constrain to an accuracy loosely characterized by the
analysis above.  

Therefore, for the power-law mass distribution, we expect the posterior distribution of (log) rate and powerlaw exponent
will be correlated and follow a Gaussian distribution characterized by  the inverse covariance 
\begin{eqnarray}
\Gamma \simeq 
\mu \begin{bmatrix}
1 & -1 \\
-1 & 1+2 \partial_\alpha^2 \ln C(\alpha)
\end{bmatrix}
\label{eq:FisherEstimate:PowerlawRate}
\end{eqnarray}
relative to the coordinates $(\ln {\cal R},\alpha)$, 
if we adopt a uniform prior on $\alpha$ and $\ln R$.  
This expression captures the correlations between the rate and mass ratio seen in our inferences, when only
varying the total event rate and mass ratio.

\subsection{Semianalytic model for constraints on  the spin magnitude and misalignment distribution}
\label{ap:sub:SpinConstraints}

In this paper, for the purposes of illustration and as a leading-order approximation suitable for the BH-BH binaries
reported to date, we adopt three simplifying approximations: that the sensitive volume depends weakly on spin;  that GW
measurements will only constrain $\chi_{\rm eff}$; and that the underlying mass and spin distributions of BH-BH binaries
are uncorrelated.  In this framework of approximations,  only  $\chi_{\rm eff}$ measurements and hence the underlying $\chi_{\rm eff}$
distribution of the population determines how well we can distinguish between population models via spin measurements.
Within this framework, we can simply and largely analytically estimate how much information we gain about the BH spin
distribution from repeated measurements.

\begin{figure}
\includegraphics[width=\columnwidth]{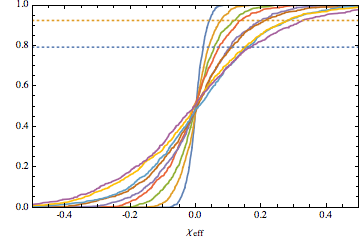}
\includegraphics[width=\columnwidth]{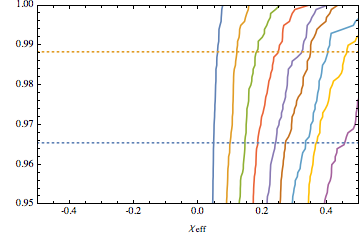}
\caption{Why $\chi_{\rm eff}$ measurements constrain the maximum spin: cumulative distribution function for $\chi_{\rm eff}$ for toy models with isotropic spins and uniform spin magnitude distributions limited
  by $0.1,0.2, 0.3, \ldots$.
In the top panel, the vertical lines, corresponding to $0.5^{1/3}$ and $0.5^{1/25}$, indicate the locus of points in each cumulative distribution function we can begin
to constrain  with the absence of events above $X$ with $3$ and 25 events, respectively.
In the bottom panel, the lines have been changed to $0.9^{1/3}$ and $0.9^{1/25}$, respectively.  
}
\label{fig:ToyModelConstraintOnSpinMagnitude}
\end{figure}

In our synthetic model (and nature) where BH spins appear to be small, the first few measurements will principally inform our
upper limit on the BH spin distribution, via the absence of observations consistent with large $\chi_{\rm eff}$.   For
example, in our synthetic model, the 90\% upper limit expected in 25 events is
$\chi_{\rm eff} < 0.31$; for our inferred posterior predictive distribution based on all
published  events, it is $0.19$.
In Fig. \ref{fig:ToyModelConstraintOnSpinMagnitude}, we use a simple toy model to illustrate how upper limits loosely
inform our estimates of the BH spin distribution.  In this model, we  assume each BH in a binary has a random spin
magnitude drawn from a  uniform distribution between  0 and $\chi_{\rm max}$,  randomly (isotropically) oriented, for
binaries with a random mass ratio uniformly drawn between $0.1$ and $1$.  
This figure shows the cumulative distribution of $\chi_{\rm eff}$ implied by these assumptions, for different choices of
$\chi_{\rm max}$.  These cumulative distributions are well approximated by analytic expressions for the cumulative
  distribution of $\chi_{1,z}$ and $\chi_{\rm eff}$ under these assumptions; see  \cite{gwastro-PENR-RIFT} for concrete expressions.  For comparison, the vertical shaded regions show the largest values of $\chi_{\rm eff}$ which have
significant support in our synthetic sample ($\chi_{\rm eff} \lesssim 0.5$), consistent with the largest plausible spins
reported for O1 and O2 events.  The lack of support for large $\chi_{\rm eff}$ in any observation to date strongly
suggests that BH spins cannot be large. 
Conversely, an observation of a binary with $\chi_{\rm eff}$ bounded below by $\epsilon$ (e.g., GW151226) implies that
a significant fraction of BH spins must be greater than of order $\epsilon$.

We emphasize that we provide these estimates (and perform our calculation within these underlying approximations) to
produce a \emph{conservative, well-understood benchmark} for how well the BH spin distribution can be constrained with
present and future GW measurements.  Real GW measurements, particularly of low-mass or closer and therefore
higher-amplitude BH-BH mergers, will provide additional direct constraints on the other spin degrees of freedom.

\section{End-to-end tests of population hyperparameter recovery:  $P$--$P$ plots}
\label{ap:pp-plots}

\begin{figure}

\includegraphics[width=\columnwidth]{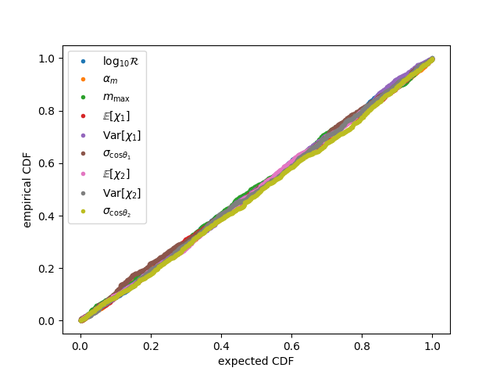}
\includegraphics[width=\columnwidth]{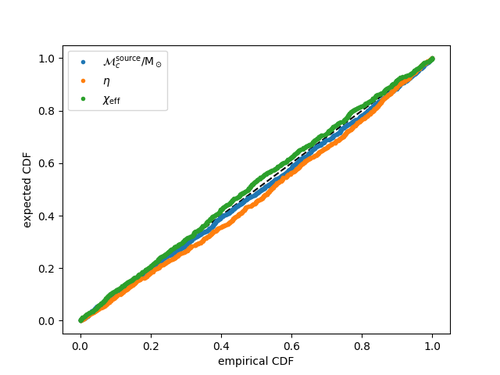}

\caption{$P$--$P$ plots for hyperparameter recovery. Top (bottom) panel shows the $P$--$P$ plot for population (single synthetic event) inferences.}
\label{fig:pp-plots}
\end{figure}

A standard technique to test Bayesian parameter inference codes is a probability-probability or $P$--$P$ plot.  We employ this test both on our population inference engine and on the procedure for making synthetic observations.
For our population inference code, we generate $k=1 \cdots 1000$ synthetic BBH populations, each a fair draw from a set of population
hyperparameters controlling the rate, mass and spin distribution.  For each synthetic population, we generate one random
observing run with O1 LIGO sensitivity and $T = 300\,\mathrm{days}$ coincident observing time, by computing the expected
number of detections $\mu$ [Eq. \ref{eq:mu-from-VT}] and taking one random Poisson draw $p(n_k) \propto e^{-\mu}
\mu^{n_k}/n_k!$.  We take $n_k$ detection-weighted binaries, generating parameter estimates according to the procedure
in Appendix \ref{ap:mockPop}.  We then apply our population parameter inference code to generate posterior distributions
on the population hyperparameters $\Param$, and from that one-dimensional marginal cumulative distributions
$\hat{P}_{k,i}(\Param_i)$, for each parameter $\Param_i$.  It should be noted here that we used as our prior the same
distribution that these population hyperparameters were drawn from, as anything else would produce biases.  Using the
true hyperparameter values $\Param^*_{k,i}$, we generate a single number for each hyperparameter
$\hat{P}_{k,i}(\Param^*_{k,i})$.  A $P$--$P$ plot is the cumulative distribution of these $\hat{P}_{k,i}(\Param^*_{k,i})$.
If the code is behaving correctly, these should be uniformly distributed from 0 to 1:  the plot should be diagonal.  
The top panel of Fig. \ref{fig:pp-plots} shows the $P$--$P$ plots for each of our model hyperparameters. %

In addition to our population inference code, we made $P$--$P$ plots for our synthetic parameter estimation code, described
in Appendix \ref{ap:mockPop}, as our population inference tests make use of it.  Here we generated $k=1 \cdots 1000$
synthetic BBH signals, drawing true values from the prior we used for measuring the posteriors.  We repeated the same
process just described, making posterior distributions on the intrinsic parameters $\param$, and evaluating the marginal
cumulative distribution functions at the true values $\param^*_k$.  $P$--$P$ plots for some representations of the intrinsic parameters are shown in the
bottom panel of Fig. \ref{fig:pp-plots}.  

\bibliography{references,paperexport,LIGO-publications}

\end{document}